# Properties of outer solar system pebbles during planetesimal formation from meteor observations

Peter Jenniskens [1,2,*], Paul R. Estrada [2], Stuart Pilorz [1], Peter S. Gural [3], Dave Samuels [1], Steve Rau [4], Timothy M. C. Abbott [5], Jim Albers [1], Scott Austin [6], Dan Avner [7], Jack W. Baggaley [8], Tim Beck [9], Solvay Blomquist [7], Mustafa Boyukata [10], Martin Breukers [4], Walt Cooney [11], Tim Cooper [12], Marcelo De Cicco [13], Hadrien Devillepoix [14], Eric Egland [1], Elize Fahl [15], Megan Gialluca [7], Bryant Grigsby [1], Toni Hanke [15], Barbara Harris [16], Steve Heathcote [5], Samantha Hemmelgarn [7], Andy Howell [17], Emmanuel Jehin [18], Carl Johannink [4], Luke Juneau [19], Erika Kisvarsanyi [20], Philip Mey [21], Nick Moskovitz [7], Mohammad Odeh [22], Brian Rachford [23], David Rollinson [24], James M. Scott [25], Martin C. Towner [13], Ozan Unsalan [26], Rynault van Wyk [14], Jeff Wood [24], James D. Wray [1], C. Pavao[1], and Dante S. Lauretta [27].

[1] SETI Institute, 339 Bernardo Ave, Mountain View, CA 94043, USA.

[2] NASA Ames Research Center, Mail Stop 245-3, Moffett Field, CA 94035, USA.

[3] Gural Software and Analysis LLC, 12241 Eliza Court, Lovettsville, VA 20180, USA.

[4] CAMS BeNeLux, Am Ollenkamp 4, D 48599 Gronau, Germany.

[5] CAMS Chile, Cerro Tololo Inter-American Observatory, NSF's National Optical-Infrared Astronomy Research Laboratory, Casilla 603, La Serena, Chile.

[6] Dep. of Physics and Astronomy, University of Central Arkansas, 201 Donaghey Ave, Conway, AR 72035, USA.

[7] LOCAMS, Lowell Observatory, 1400 West Mars Hill Road, Flagstaff, AZ 86001, USA.

[8] CAMS New Zealand, Dept. of Physics & Astronomy, University of Canterbury, Christchurch 8140, New Zealand.

[9] Mendocino College, 1000 Hensley Creek Road, Ukiah, CA 95482, USA.

[10] CAMS Turkey, Yozgat Bozok University, Department of Physics, 66100, Yozgat, Turkey.

[11] CAMS Texas, 4635 Shadow Grass Dr., Katy, TX 77493, USA.

[12] CAMS South Africa, Suite 617, Private Bag X043, Benoni 1500, South Africa.

[13] CAMS EXOSS, Observatorio Nacional, Rua Gal. José Cristino 77, Rio de Janeiro, RJ 20921-400, Brazil.

[14] CAMS Australia, International Centre for Radio Astronomy Research, Curtin University, Perth WA 6102, Australia.

[15] CAMS Namibia, High Energy Stereoscopic System Experiment, Windhoek, Namibia 11009.

[16] CAMS Florida, BarJ Observatory, New Smyrna Beach, FL 32169, USA.

[17] CAMS Florida, Gainesville, FL 32605, USA.

[18] CAMS Chile, STAR Institute, University of Liège, B-4000 Liège 1, Belgium.

[19] CAMS Arkansas, North Little Rock, AR 72118, USA.

[20] College of Central Florida, 3001 SW College Road, Ocala, FL 34474-4415, USA.

[21] South African Radio Astronomy Observatory, HartRAO, Farm 502 JQ, Broederstroom Road, Hartebeesthoek, 1740, South Africa.

[22] UACN, International Astronomical Center, P.O. Box 224, Abu Dhabi, UAE.

[23] Embry-Riddle Aeronautical University, 3700 Willow Creek Road, Prescott, AZ 86301, USA.

[24] P.O. Box 162, Willetton, WA 6955, Australia.

[25] Department of Geology, University of Otago, P. O. Box 56, Dunedin 9054, New Zealand.

[26] CAMS Turkey, Ege University, Faculty of Science, Department of Physics, 35100, Bornova, Izmir, Turkey.

[27] Lunar and Planetary Laboratory, University of Arizona, Tucson, AZ 85721, USA.

*) Corresponding author. Email: petrus.m.jenniskens@nasa.gov

**Abstract.**

Observations of proto-planetary disks, as well as theoretical modeling, suggest that in the late stages of accretion leading up to the formation of planetesimals, particles grew to pebbles the size of 1-mm to tens of cm, depending on the location and ambient conditions in the disk. That is the same size range that dominates the present-day comet and primitive asteroid mass loss. Meteoroids that size cause visible meteors on Earth. Here, we hypothesize that the size distribution and the physical and chemical properties of young meteoroid streams still contain information about the conditions in the solar nebula during these late stages of accretion towards planetesimal formation. If so, they constrain where long-period (Oort Cloud) comets, Jupiter-family (Scattered Disk Kuiper Belt) comets, and primitive asteroids (Asteroid Belt) formed. From video and visual observations of 47 young meteor showers, we find that freshly ejected meteoroids from long-period comets tend to have low bulk density and are distributed with equal surface area per log-mass interval (magnitude distribution index $\chi \sim 1.85$), suggesting gentle accretion conditions. Jupiter-family comets, on the other hand, mostly produce meteoroids twice as dense and distributed with a steeper $\chi \sim 2.15$ or even $\chi \sim 2.5$, which implies that those pebbles grew from particles fragmenting in a collisional cascade or by catastrophic collisions, respectively. Some primitive asteroids show $\chi >$ 2.5, with most mass in small particles, indicating an even more aggressive fragmentation by processes other than mutual collisions. Both comet populations contain an admixture of compact materials that are sometimes sodium-poor, but Jupiter-family comets show a higher percentage (~8% on average) than long-period comet showers (~4%) and a wider range of percentages among comets. While there are exceptions in both groups, the implication is that most long-period comets formed under gentle particle growth conditions, possibly near the 30 AU edge of the Trans Neptunian Disk, while most Jupiter family comets formed closer to the Sun where pebbles reached or passed the fragmentation barrier, and primitive asteroids formed in the region where the cores of the giant planets formed. This is possible if the Scattered Disk represents all objects scattered by Neptune during its migration, while the present-day outer Oort cloud formed only during and after the time of the planet instability, well after the Sun had moved away from sibling stars.

## 1. Introduction

It is widely accepted that proto-comets and primitive asteroids must have formed from a settled mm- to dm-sized pebble layer with sufficient solids-to-gas ratio allowing self-gravitating over-densities to develop that then collapsed into larger bodies tens of kilometers in diameter (see reviews by Weissman et al., 2020; Blum et al., 2022). Millimeter observations with the Atacama Large Millimeter/submillimeter Array (ALMA) (e.g., Andrews et al., 2018; Long et al., 2018) and the Very Large Array (e.g., Wilner et al., 2018) have observed pebbles in this size range across protoplanetary disks everywhere. This leapfrogging from an array of pebbles to km-scale comes about because particle growth is inhibited due to a variety of growth barriers, especially in turbulence (Johansen et al., 2007; Rosotti, 2023).

Primary accretion, the stage of growth from submicron size dust grains to planetesimals, remains the least well-understood stage of the evolution of solids in the proto-planetary nebula because, even under weakly turbulent conditions, growth stalls at relatively small sizes, indicating that incremental growth is not a viable pathway to planetary building blocks. Yet, primitive chondritic parent bodies apparently did accrete, as early as ~0.5 Myr, with this process continuing over a period of several million years (e.g., Connelly et al. 2012; Kruijer et al., 2017; Schrader et al., 2017).

Central to the problem of continued growth are the various barriers pebbles may encounter when they decouple from the gas and settle towards the disk midplane. These include the bouncing (e.g., Windmark et al., 2012) and fragmentation barriers (e.g., Dullemond & Dominik, 2005; Johansen et al., 2016; Tominaga & Tanaka, 2023), which generally frustrate further growth beyond mm–cm sizes. Also important is the radial drift barrier, in which decoupled pebbles drift inward due to gas drag faster than they can grow (Brauer et al., 2008; Birnstiel et al., 2010). Estrada et al. (2016) explored global models of solids evolution and redistribution of condensables in an early turbulent, viscously evolving protoplanetary nebula, in which they self-consistently treated the growth and radial drift of solid particles of all sizes and found that growth beyond ~mm-cm sizes is limited by a combination of these barrier mechanisms consistent with the findings of other workers (e.g., Brauer et al., 2008; Birnstiel et al., 2009; Zsom et al., 2010). This particular model indicated that particles grow into pebbles with diameters of order 20 cm at 4 AU to only 1 mm at 40 AU from the Sun. More recent models show that fractal growth into low-density materials outside the snowline allows particles to remain coupled to the gas longer, but their enhanced cross sections eventually lead to accelerated growth and rapid inward drift from the snow line out to ~20 AU (Estrada et al., 2022).

These barriers can be mitigated if the nebula is not turbulent (e.g., Weidenschilling, 2011), but this is increasingly inconsistent with observations in which some level of mixing is indicated, not only from the observed admixture of compact (high-density) materials in comets (e.g., Hörz et al., 2006; Levasseur-Regourd et al., 2018) and the wide range in the ages of primitive meteorites but also from the fact that even meteorites of the same age are compositionally distinct and were formed spatially separated (e.g., Wadhwa & Russell, 2000; Cuzzi & Weidenschilling, 2006; Simon et al., 2018). This view is supported observationally by both direct estimates of turbulence through line-broadening in proto-planetary disks (e.g., Flaherty et al. 2018), as well as estimates using indirect methods (for a review, see Rosotti, 2023).

The difficulties associated with continued growth has led to the argument that some mechanism acts to leapfrog from growth-frustrated pebbles to tens-of-km-sized bodies if those mm-to-cm-sized pebbles can settle or be concentrated into a sufficiently dense layer. This collapse into planetesimals (comets and the first asteroids) could happen during the time period when the gas disk gradually dissipates and the dust-to-gas ratio increases (e.g., Youdin & Goodman, 2005; Hartlep & Cuzzi, 2020).

Here, we hypothesize that the size distribution and some physical properties of the cm-sized meteoroids (composed of grains) as they are ejected from comets (Jenniskens, 2006; 2016) and primitive asteroids (Lauretta et al., 2019) still reflect those of the pebbles (composed of particles) in the proto-planetary nebula at the time of collapse.

Comets are weak assemblages of these pebbles, accumulated in a low-gravity environment. The fine-grained dust captured by the Stardust mission to comet 81P/Wild 2 (e.g., Brownlee et al., 2006; 2012; Hörz et al., 2006) and the material studied by the Rosetta Mission to comet 67P/Churyumov-Gerasimenko (e.g., Bentley et al., 2016; Fulle et al., 2018; Levasseur-Regourd et al., 2018) both showed that these meteoroids are a conglomeration of larger inner-solar system solids, fine-grained outer-solar system materials containing organics, and various ices, with materials assembled both as low-density ("fluffy") meteoroids and high-density meteoroids. While ices add to the volume, the mineral and organic components determine much of the cohesion of the meteoroids. When meteoroids are ejected, the volatile ices evaporate, leaving the organic and mineral matter as meteoroids.

Until now, direct measurements of the size distribution of solids in the mm–cm size domain are constrained only by that of chondrules in chondritic meteorites, if it is assumed that the chondrules represent nebula solids and are not derived from impact melts (Ormel et al., 2007).

Meteoroids of this size would cause visible meteors on Earth (e.g., Jenniskens, 2006). The ejecta from a given comet evolves into a meteoroid stream that can produce a meteor shower when the stream intersects Earth's path. All meteoroids in a stream approach Earth's atmosphere at close to the same speed, so that the distribution of kinetic energy (luminosity) reflects the distribution of meteoroid mass. Each individual meteoroid will disrupt in the atmosphere in a manner reflecting how it is built up of smaller grains and the meteor beginning altitude and deceleration depend on the meteoroid density. The emitted light can be probed spectroscopically to determine the meteoroid main-element composition (Ceplecha et al., 1998; Borovicka, 1998; Trigo-Rodriguez et al., 2003; Jenniskens, 2007).

This makes meteor observations a powerful tool, short of comet sample return, for measuring the mass distributions of cometary ejecta at cm size scales (Trigo-Rodriguez & Blum, 2022), as well as the mass distribution of constituent grains that make up the meteoroids (Beech & Murray, 2003). In particular, photographic, visual and video techniques detect the large meteoroids in the relevant 1 mm to 10 cm-size range that cause visible $m = +5$ to $-12$ magnitude meteors (Borovicka et al., 2022; Jenniskens, 2023, Chapt. 3).

The Cameras for Allsky Meteor Surveillance (CAMS) project developed video technologies to scale up the triangulation of visible meteors (Jenniskens et al., 2011) and has now measured over 2.7 million atmospheric trajectories and pre-atmospheric orbits from 2010 to 2023. 29% of those meteors belong to 487 meteor showers and shower components from perhaps 200 individual parent bodies (Jenniskens et al., 2016a–c; 2018; Jenniskens & Nénon, 2016; Jenniskens, 2023).

In previous work, we derived fundamental parameters for 513 meteor showers (Jenniskens, 2023). The data was validated against previous studies of a small number of major showers in the literature (e.g., Beech & Murray, 2003; Brosch et al., 2004; Blaauw et al. 2016; Buccongello et al., 2023). 26 of those showers were detected uniquely by radar. This raw data was published in the book "*Atlas of Earth's Meteor Showers*" (Jenniskens, 2023). Subsequently, we examined from this data how meteoroid streams evolve over time, deriving dependencies on perihelion distance and inclination of the orbit (Jenniskens et al., 2024). That work focused on evolved showers and streams in orbits close to the Sun.

Here, we focus on the dynamically youngest meteor showers, derive new data for the youngest shower components, and use these results to investigate in what physical form cm-sized meteoroids are ejected from different dynamical classes of comets. The results are discussed in the context of local conditions in the solar nebula at the time of the collapse of pebbles into planetesimals in the regions that produced the Oort Cloud, the Kuiper Belt, and the Asteroid Belt.

## 2. Methods

The relevant meteoroid properties derived from meteor observations include mass, the differential mass frequency distribution for the stream, the grain mass distribution, the density, beginning altitudes of ablation, and the sodium (Na), magnesium (Mg) and iron (Fe) elemental composition. Methods on how these properites are derived from low-light video observations of meteors are described in Jenniskens (2023, Chapter 2) and Jenniskens et al. (2024). Here, we briefly summarize the approaches and add details relevant to this work.

### 2.1. Mass (m) and mass frequency distribution (s) from meteor magnitude ($M_v$) and magnitude distribution ($\chi$)

In a given meteor shower, the absolute magnitude of meteors (the irradiance when seen from a distance of 100 km) is proportional to kinetic energy, and because all meteoroids have close to the same entry speed in a shower, that means that the meteor magnitude frequency distribution is proportional to the meteoroid mass frequency distribution. Meteor brightness is expressed as magnitudes $M_v$ in the astronomical V photometric passband centered on 548 nm wavelength. With radar, video and photographic techniques taken together, meteor observations cover a wide range of masses (m), from about m = 0.0001g ($M_v$ = +10 magnitude) to 10 kg (-10 magnitude). Low-light video observations, from which our data is derived, mostly detect meteors in the range $M_v$ = +5 to -5 magnitude (mass 0.01 to 100 g).

The mass of a zero-magnitude meteor, the typical brightness of a video-detected meteor at the peak of the observed distribution, is different for different meteor showers. The brightness is proportional to kinetic energy, so that the mass of a zero-magnitude meteor is proportional to the inverse square of the apparent entry speed ($V_\infty$).

When translating from mass to diameter, the size of a meteoroid of given mass depends on the density of the meteoroid. The density of the meteoroid is derived from how quickly the meteoroid slows down on entering Earth's atmosphere and at what altitude that occurs (Jenniskens et al., 2024).

Figure 1 shows the median diameter of a meteoroid causing a +0 magnitude CAMS-detected meteor for all meteor showers studied here. The median diameter changes from 0.7 cm at $V_\infty = 72$ km/s for long-period comet showers to about 1.8 cm for $V_\infty = 11$ km/s more typical of Jupiter-family type comets and primitive asteroids, which both are in the relevant range of pebbles formed in the protoplanetary disk.

Dashed lines in Fig. 1 show the corresponding range of meteoroid sizes for magnitudes in the range $M_v = +5$ to -5. Within that range of magnitudes, the meteor's peak brightness along its path is distributed according to $N(M_v) \sim \chi^{Mv}$, with $\chi$ the magnitude distribution index. The observed distributions need to first be corrected for the magnitude-dependent detection probability $P(M_v)$, with the observed distribution $N(M_v) = P(M_v)\ \chi^{Mv}$ peaking around +0 magnitude. Those probabilities depend on the meteor shower entry speed and the type of camera used. Details are given in Jenniskens (2023, Chapt. 2).

Typical values for the magnitude distribution index range from $\chi \sim 1.5$ to 4.7. Figure 2 shows examples of corrected magnitude distributions for two observed showers with a total number N = 100 meteors. There is little evidence that the distribution deviates from a simple power law over that magnitude range (Jenniskens, 2006; 2023).

A meteoroid stream does not contain a "typical" meteoroid size, but rather a range of sizes. The mass-dominant size is either towards the high mass end of the observed distribution if $\chi \ll 2.52$, or towards the low end if $\chi \gg 2.52$. Most streams have no known high-mass or low-mass cut-off to the size frequency distribution within the range of observed masses, but there is likely a lack of masses larger than 10 kg, because asteroids, not comets, dominate the influx of masses >10 kg. There may also be a cut-off at masses < 0.0001 g because few streams register as features in the zodiacal cloud (Jenniskens, 2006). If the mass of a zero-magnitude meteor is 1g, then over the +10 to -10 magnitude interval the mean mass at $\chi = 1.5$ is 7.5 kg (diameter d ~ 24 cm), at $\chi = 2.5$ is 0.002 g (d ~ 0.15 cm) and the mean mass at $\chi = 4.7$ is 0.002 g (d ~ 0.07 cm).

All meteors in a shower enter Earth's atmosphere at close to the same apparent entry speed. In that case, kinetic energy is proportional to mass and the differential mass distribution index s = 1 + 2.5 log ($\chi$). Values range from s = 1.4 to 2.7. The differential size frequency power-law index ($\alpha$) follows, assuming the meteoroid density is constant: $\alpha = 3s - 2$. Values of $\alpha$ range from $\alpha$ = 2.3 to 6.0. Sporadic (non-shower) meteors of +4 to -2 magnitude have $\chi \sim 3.4$ (Jenniskens, 2006).

## 2.2 Meteoroid composite grain frequency distribution from meteor lightcurve shape parameter (F) and grain magnitude distribution ($\chi_g$)

The meteor light curve shape is a unique diagnostic of the mass frequency distribution of the grains that compose a meteoroid. A non-fragmenting single-grain meteoroid would have a classical light curve that has a sudden onset when it gets hot enough to ablate (T ~ 2000 K), then an exponential increase in brightness in response to the exponentially increasing air density, followed by a late peak, a turnover, and finally a rapid drop-off when enough material has ablated for the meteoroid surface area to decline (Öpik, 1958; Beech, 2004). Most meteors show a more symmetric light curve shape due to fragmentation (Jacchia, 1955).

In the "dust-ball" model (Hawkes & Jones, 1975), the meteoroid is assumed to have fallen apart into individual components before reaching temperatures high enough to ablate individual grains. The meteor light curve is then a sum of classical light curves for individual grains. This model has since been developed further, describing the physical processes of heating and ablation of the material together with a description of the fragmentation process (Campbell-Brown & Koschny, 2004; Campbell-Brown et al., 2013; Borovicka et al., 2007; Beech, 2009; Pecina & Koten, 2009). The models fit both light-curve and velocity deceleration data simultaneously. Fit parameters include uniquely constrained parameters such as grain mass distribution index (shape of the light curve), meteoroid and grain density (deceleration), and heat of ablation (beginning altitude of ablation), and not so well constrained parameters, that include temperature of "glue" release, boiling temperature, thermal conductivity, the specific heat of the meteoroid, and moments of non-regular fragmentation.

After implementing the Campbell-Brown & Koschny (2004) model to the CAMS observations, we found that the models are not well constrained by the observation of individual meteors and require the manual identification of fragmentation points to match the light curves well.

Here, we adopted the simpler approach by Beech & Murray (2003) of isolating the grain mass distribution index parameter from measuring the light curve shape parameter (F), which is defined as the position of the peak brightness ($H_m$) along the meteor altitude range of beginning height ($H_b$) and end height ($H_e$):

$$F = (H_m - H_b) / (H_e - H_b) \quad (1)$$

This light curve shape parameter (F) relates to the grain magnitude distribution index $\chi_g$, and corresponding grain mass distribution index $s_g$. Model light curves for ~0.16 cm meteoroids (Beech & Murray, 2003) show a peak early (high grain mass distribution index $s_g$) or late along the trajectory, with F = 0.40 for $\chi_g = 1.82$ ($s_g = 1.65$), F = 0.31 for $\chi_g = 2.15$ ($s_g = 1.83$) and F = 0.26 for $\chi_g = 2.51$ ($s_g = 2.00$). Leonid shower light curves were best described having derived from a low $s_g = 1.6 \pm 0.1$.

**2.3 Meteoroid density (ρ) and admixture fraction of low beginning-height materials ($f_{lb}$)**

Meteoroid bulk densities were derived from the measured decelerations and the magnitude +0 meteor beginning heights (see details in Jenniskens et al., 2024, Sect. 2.2). The median results correlate well with densities derived by Buccongello et al. (2023) based on high-spatial and high-temporal resolution imaging of meteors.

It was found that each shower consists of a group of meteors that has high beginning heights (height class Type I) and a group with low beginning heights (Type III). The height class was defined by correcting the beginning altitude for a $V^2$-dependence and calculating a height class parameter ($k_c$) in the range 85–105 km after (Ceplecha 1967, 1988; Ceplecha & McCrosky, 1976; Jenniskens et al., 2016c). By grouping $k_c$ into 3-km intervals, Jenniskens (2023) classes Ia to II form an upper band (called Type I), while classes IIb to IV form a lower band (called Type III) (Jenniskens, 2023, Fig. 2.5).

Figure 3 shows the example of shower #842. The meteoroid density for the high height-class group tends to come out slightly lower than those for the low height-class group. Within the scatter from measurement errors, the meteoroid densities of individual meteors in this shower do not strongly correlate with height in each height class (Fig. 3). In this paper, when we discuss density (ρ), we refer to the values for the Type I group.

There are several reasons why some meteoroids would start ablating higher in the atmosphere than others, including differences in strength, in thermal conductivity, and in composition, with high beginning heights expected for materials that are fragile, have low thermal conductivity, and contain more volatile components. Our observations suggest that these properties are related, and that meteor showers contain two types of meteoroids. From the scattering properties of cometary dust and from in-situ observations of comet dust, cometary matter is known to contain both low density ("fluffy") and high density ("solid") materials (e.g., Levaseur-Regourd et al., 2018).

The admixture fraction $f_{Ib}$ equals the number of detected meteors III/(I+III). This is not a mass fraction, because the magnitude distributions for Type I and III materials may be different. In the case of the Lyrid meteor shower (IAU#6), for example, $\chi = 2.18 \pm 0.05$ for Type I and $\chi = 2.70 \pm 0.20$ for Type III.

**2.4. Meteoroid elemental composition from meteor spectroscopy**

The different beginning heights could point to meteoroids (or the original pebbles) having been heated in the past to the point of changing their physical properties. Meteor spectroscopy measures changes in elemental compositions, a technique similar to laser-induced breakdown spectroscopy (LIBS) used to measure the elemental composition of meteorites.

A 16-camera CAMS Spectrograph (Fig. 4A) has been in operation in the coverage area of CAMS California since March of 2013. The spectrograph consists of 16 *Watec Wat902H2 Ultimate* cameras equipped with a 1379 grooves/mm fused silica grating (*Ipsen Photonics*) and 12mm f1.2 Pentax lens, spreading the light over 0.6 nm/pixel in 1$^{st}$ order and 0.3 nm/pixel in 2$^{nd}$ order. Since first light, the CAMS Spectrograph has collected thousands of meteor spectra for the brightest meteors (≤ 0 magnitude) for which the meteor beginning heights, orbital elements and shower identification are known (Jenniskens, 2023, p. 28-29).

A total of 1005 of these spectra were reduced, recorded during the period March 2013 to August 2014. Data reduction included first an identification of the meteor spectra in the video data, for which a deep learning tool was developed. The next step was to identify spectra that are spatially and temporally coincident with the CAMS meteor trajectories. For that, a software *SP_Coincidence* was written (Fig. 4B, and caption). The known positions of stars (most outside the field of view) were used to calculate the position of star spectra in each objective grating camera, and to extract the star spectra for calibrating the instrument responsivity and atmospheric extinction. The known position of the meteor in the field of view of the spectrograph was then used to extract the meteor spectrum. To do so, the observer adjusts the parameters of yaw, roll and pitch of the grating, as well as the grating dispersion (lines/mm) to optimize extraction of the 1-dimensional spectrum. The 1-dimensional spectrum was integrated along the observed meteor trajectory.

The meteor's emitted light is dominated by atom line emissions, in particular from meteoric metals magnesium (Mg), iron (Fe) and sodium (Na), and from atmospheric oxygen (O) and nitrogen (N) atoms generated by the meteor. Slow meteors emit mainly sodium and graybody continuum emission, while fast meteors also emit light in the atmospheric First Positive Band of $N_2$ (e.g., Matlovic et al., 2020; Jenniskens, 2023, Fig. 3.5).

Here, we consider the integrated line intensities along the whole meteor trajectory only. Some meteoroids ablate a significant fraction of sodium before losing most magnesium and iron. This has been interpreted as evidence that such meteoroids are more porous, exposing the low boiling point sodium-containing minerals before ablating the magnesium and iron. This phenomenon is not common, but young Leonid showers are known for this (e.g., Beech & Murray, 2003; Campbell-Brown et al., 2000).

The early solar system chondritic abundances relative to Mg are: Mg = 1.00, Fe = 0.85 and Na = 0.056 (Lodders, 2019). These abundance ratios translate into line intensity ratios of Mg, Fe and Na in an entry-speed ($V_\infty$) dependent way (e.g., Trigo-Rodriguez et al., 2003; Jenniskens, 2007). Figure 5A shows the measured line intensities compared to those of oxygen, after correcting for instrument responsivity and atmospheric extinction. The ratio of Na/O, Fe/O and Mg/O are shown as a function of meteor entry speed. Most values cluster along a narrow band, which we assume are the relative line intensities expected for normal chondritic elemental abundances. The best-fit trend of those line intensities are (dashed lines in Fig. 5A):

$$\text{Mg/O} = 0.10 + 4.3 * exp(-0.06 * V_\infty) \quad (2)$$

$$\text{Fe/O} = 0.19 + 15.2 * exp(-0.11 * V_\infty) \quad (3)$$

$$\text{Na/O} = 0.077 + 110 * exp(-0.17 * V_\infty) \quad (4)$$

Relative abundances can be expressed as the ratio of the measured line-intensity-ratio to that expressed by Eq. 2–4, multiplied by the chondritic abundances of Lodders (2019). Those values are given in Jenniskens (2023).

The line intensity ratios of the meteoric metals Fe/Mg and Na/Mg are shown as a function of $V_\infty$ in Fig. 5B. While the scatter in Fe/Mg is mostly due to measurement error, a significant portion of meteors have lower-than-normal Na/Mg line ratios.

## 2.5. Meteoroid stream evolution and meteoroid thermal history

When meteoroid streams are first created, the meteoroids move in nearly parallel orbits (which would produce nearly the same meteor radiant) and with nearly equal velocity relative to Earth. Only a small number of meteor showers are known that have the compact radiant and entry speed distributions that result from the initial ejection and spreading along the comet orbit, before streams are more dispersed by planetary perturbations. Their physical parameters and magnitude distribution index are evaluated here directly without correction for age.

For the bulk of showers, however, it is important to understand how the physical parameters and magnitude distribution index were changed in the interplanetary medium since ejection from the comet. The current orbit of the meteoroids (with the perihelion distance being the most relevant for addressing the issue of heating) may not have been the same as that in the past.

This was investigated in an earlier study (Jenniskens et al., 2024). Meteoroid stream ages for short-period comet showers were taken from the literature, while new long-period comet shower ages were calculated by the dynamical modeling of long-period comet meteoroid streams. Methods followed Vaubaillon et al. (2005), Jenniskens (2006), Jenniskens & Vaubaillon (2010), Moorhead et al. (2015), and Abedin et al. (2017), taking into account the planetary perturbations of all the planets from Mercury to Neptune with added modules for radiation pressure and the Poynting-Robertson effect. The approach was further described in Jenniskens et al. (2024).

It was found that long-period comet showers disperse linearly with age for the period 10,000 to 120,000 years (after an initial rapid increase of dispersion in the first few orbits when material first spreads around the comet orbit). The reason for this is a combination of a root-of-N dependence from inner solar system perturbations (with N the number of times that the meteoroid passes through the inner solar system), combined with a dispersion of orbital periods, whereby most orbits that frequent Earth evolved to small P ~ 250 y orbits. The measured dispersion at Earth's orbit was then used to calculate the age of the shower (Jenniskens et al., 2024).

It was found that the magnitude distribution index increases over time, with large meteoroids being preferentially removed from the population compared to small meteoroids in the −5 to +5 magnitude range. The rate of that evolution depends on the square root of the perihelion distance (q) of the orbit, suggesting that peak temperature is the determining factor that limits the lifetime of these meteoroids (Jenniskens et al., 2024). The age of a meteor shower as derived from the shower's magnitude distribution index (Age-$\chi$) was approximately given by:

$$\text{Age-}\chi \text{ (y)} = 40{,}000 \ (\pm 8{,}000) \sqrt{q} \ (\chi - 1.85) \qquad (5)$$

Adopting the age derived from the dispersion of the stream, we used this equation to translate the observed magnitude distribution index to the initial index at the time of ejection.

## 3. Results

### 3.1. Shower source regions and dynamical evolution

We are interested here in those showers that have a known source region (Oort Cloud, Kuiper Belt, or Asteroid Belt) and that did not evolve to orbits with perihelion distance < 0.3 AU over the age of the stream, so that the meteoroids themselves were not heated to the point of losing organic matter (Jenniskens et al., 2024). The meteor showers originate from three distinct dynamical groups of minor bodies in the solar system: long-period comets, Jupiter-family comets and asteroids.

*Long-period comets (LPC)*: About half of the showers (248 out of 513) originated from long-period comets. These showers have median Tisserand Parameters with respect to Jupiter $T_J < +0.63$ and median semi-major axis > 5 AU (Jenniskens et al., 2024). Their parent bodies have orbital periods in the range 250 to 4000 years that cause meteoroid streams not too diluted to be detected (Jenniskens et al., 2021). The long-period comets that produce our meteor showers have perihelion distance q < 1.05 AU, but they originated from much higher q-orbits and aphelia in the outer reaches of the Oort Cloud (Oort, 1950), where galactic tides and stellar encounters changed the perihelion distance back into the planetary system. That region is replenished from the inner Oort Cloud (Hills, 1981) and LPCs possibly sample also the inner Oort Cloud (Kaib & Quinn, 2009).

The rare *Halley-type comets* (HTC), such as 1P/Halley and 55P/Tempel-Tuttle, evolved from long-period comets (Nesvorny et al., 2017). They fall in the long-period comet group here, even though their parent bodies have a shorter orbital period (20–250 y). That is because the detected long-period comet meteoroids also evolve their orbits on average to lower semi-major axis over time. Mean-motion resonances complicate the dynamical evolution of parent comets, but most of the HTC showers are prominent showers and their dynamical evolution was studied in detail (e.g., references in Jenniskens, 2023). None evolved to orbits q < 0.3 AU over the lifetime of the streams.

Another 83 showers are Mellish-type showers ($+0.63 \leq T_J < +1.7$) in prograde high-inclined orbits. We will only consider the long-period comet group with q > 0.3 AU here, because some members in the Mellish group can evolve to perihelion distances q < 0.3 AU due to Kozai resonances and their orbital evolution has not been investigated before (see Section 4.1).

*Jupiter-family comets (JFC)*: 69 showers are from Jupiter-family comets. The showers have median values $+2.0 \leq T_J < +3.2$ and prograde orbits. While most showers have inclinations i < 20º, a relatively high number have higher inclinations 20º < i < 80º, in part because high-inclination showers are easier to separate from the sporadic background. We will consider only the young showers with q > 0.3 AU that have not evolved their perihelion distance to q < 0.3 AU since ejection. While these objects now have short ~6-year orbital periods after capture by Jupiter, they originated from the Scattered Disk population of the Kuiper Belt, which through close encounters with Neptune feeds a steady population of JFC inward (Kuiper, 1951; Jewitt & Luu, 1993; Tancredi, 1995; Duncan & Levison, 1997; Jewitt, 2008).

In addition, 22 showers or shower components are related to the Taurid Complex of 2P/Encke. While 2P/Encke is likely a Jupiter-family comet, the perihelion distance of this comet is close to the q = 0.3 AU limit and this shower is not included. Another 10 showers are on similar Encke-like orbits, but those are mostly detected by radar and the semi-major axis may have evolved to lower values over time, or is being underestimated. Those showers are also not considered here.

*Primitive asteroids (AST)*: Primitive asteroid 3200 (Phaethon), with q = 0.14 AU, is the parent body of the strong Geminid meteor shower and appears to have had a number of disruptive events in the past. 13 showers are possibly related to the Geminids, but none of these are young. Only 5 additional showers may have an origin at primitive asteroids, two are poorly observed. Only one of the four asteroidal showers listed in Table 1 has q > 0.3 AU, the 62-Andromedids (#924). This is a reflection of the Sun's importance in generating these showers. Primitive asteroids in near-Earth orbits originated in the main asteroid belt, where effects of absorption and emission of light can change the semi-major axis of asteroids enough to move them into resonances that rapidly change their perihelion distance into crossing Earth's orbit.

*Unknown:* The remaining showers have an unknown source population affiliation. 12 showers belong to 96P/Machholz, which may have either an LPC or JFC origin. 96P/Machholz is on a high inclination JFC-like orbit, but the showers evolve smoothly in Tisserand space from the Mellish-type comet shower population. 49 showers are evolved toroidal and sunskirter showers with unknown origin. They typically have short orbital periods, well decoupled from Jupiter, but could have originated from Jupiter-family comets, asteroids, or long-period comets. The two remaining showers appear to originate from sunskirting Mellish-type showers (LPC group) but have short orbital period (Jenniskens et al., 2024).

### 3.2. Properties of young meteor showers

Table 1 isolates the youngest showers from Jenniskens (2023): showers that are episodic (so little dispersed that they have highly variable peak rates each year), annual showers with a compact radiant and velocity distribution, and meteor outbursts (young shower components) of otherwise older annual showers. Adding to data given in Jenniskens (2023), Table 1 gives the key parameters considered here for each of these showers: q, the age of the young shower component since ejection, and the properties of those young showers: $\chi$, F, and $f_{ld}$. We included showers that were documented only by visual observations and that provide $\chi$ from meteor brightness distributions, but no detailed information from the meteor light curve, deceleration, or spectroscopy.

Footnotes to the table summarize the source of the data and particulars of each shower. A "1-revolution dust trail" refers to material ejected during the previous perihelion passage of the comet (Kondrat'eva & Reznikov, 1985), which has now dispersed in the comet orbit from small differences in orbital period. Smaller meteoroids tend to be ejected with higher velocity and therefore evolve more rapidly away from the comet position (Jenniskens, 2006, p. 258). The Leonid showers from comet 55P/Tempel-Tuttle, for example, had measured magnitude distribution index ranging from $\chi = 1.8 \pm 0.1$ for the 1767 ejecta seen in 2001 to $\chi = 3.05$ for the 1866 ejecta seen in 2002 (McNaught & Asher, 1998; Jenniskens, 2006, p. 628). When the parent body is in or near a mean-motion resonance with Jupiter (corresponding to a specific orbital period), meteoroids tend to get trapped for a period of time in that resonance as well, which will keep them close to the parent comet. The name "Filament" refers to such streams, and those may be most representative of material ejected by the comet. One example is the day-long Leonid outburst of bright meteors in 1998, the year that 55P/Tempel-Tuttle returned to perihelion. The showers that are known to be in mean-motion resonances are labeled by an "*" in Table 1.

We find here that young stream components rich in large meteoroids are commonly found not only in the case of HTC showers, but also among nearly all long-period comet showers (Table 1). The one exception is the γ-Piscis Austrinids (shower #1034), for which all returns we know of were rich in small meteoroids. In contrast, very few Jupiter-family comet showers are rich in bright meteors. Even in that case, episodic activity from components trapped in mean-motion resonances are common. Some of these showers are even younger than the young showers of long-period comets. There doesn't seem to be a significant difference in age, although these showers do come back to perihelion more often.

### 3.3. Admixture fraction of type III materials and Na-poor/free meteors

The spectroscopic investigation provides for the first time a large number of relative Mg, Fe and Na abundances for meteors on known orbits as a random sample throughout the year. Aside from the fact that the meteor spectrum depends on the entry speed, the two most notable results are that most meteors have similar line intensity ratios forming a band of points in Fig. 5A. That is because most cometary meteoroids have cosmic abundances of Mg, Fe and Na. The second most striking feature is that some meteors have Na/O a factor of 3–10 lower than this band, so called "Na-poor", or even a factor of > 10 lower in which case no sodium could be detected at all ("Na-free").

In prior work, three populations of Na-poor meteoroids have been identified from the relative intensity of Na vs. Mg emissions (Borovicka et al., 2005). Those included meteoroids that approached the Sun, meteoroids that were in short asteroidal orbits and rich in iron and, quite unexpectedly, meteoroids that moved in some long-period (Oort Cloud) comet orbits. One case is well established but atypical: lack of sodium is a characteristic of the $\alpha$-Monocerotids (Jenniskens et al., 1997). They also have high fraction of Type III materials.

Here, we investigated how the sodium abundance is related to the low beginning height seen in many meteors. Figure 6 shows the height class parameter $k_c$ as a function of apparent entry speed ($V_\infty$) for all 1005 meteors for which spectra were obtained. Even in this small sample of single meteors, the high (marked "I") and low (marked "III") type height classes are apparent.

Both Jupiter-family and long-period comet showers contain meteors with both high and low beginning heights. Meteors from Jupiter-family comets arrive mostly from the anthelion source direction with apparent entry speeds in the range $V_\infty = 11 – 35$ km/s, while those from long-period comets mostly come from the apex direction with $V_\infty > 50$ km/s. Long-period comet showers have slightly higher $k_c$ (and a wider range) than Jupiter family comet showers. Showers with perihelion distance $q < 0.3$ AU are marked as open circles in Fig. 6. These are mostly found in the mid-range of apparent entry speed, between $V_\infty$ = 35 and 50 km/s. They tend to have somewhat lower beginning heights in both bands.

Figure 7A compares the sodium content to height $k_c$ for all meteors with spectra. The normal chondritic Na/O ratio of ~ 0.06 mostly has high $k_c$, corresponding to the type I band. Those with orbits $q < 0.3$ AU are shown as open circles. Many low-q showers have low sodium abundances, as noticed before by Borovicka et al. (2005). They also tend to have low beginning heights. This can also be seen in the data by Matlovic et al. (2019), who has most sodium-poor meteors in the low-q group (35–50 km/s entry speed range).

Not all low-q meteors have low beginning heights. Some plot among the high beginning height group with normal sodium abundances. These are mostly showers in a transition zone $0.2 < q < 0.3$ AU, suggesting that these meteoroids did not yet evolve to low enough q in the past.

Looking at meteoroids with orbits $q > 0.3$ AU, Figure 7B shows that same data. Open circles isolate the meteors from the apex source that originated from long-period comets. These comets never had orbits $q < 0.3$ AU since ejection of the meteoroids (see below). They nearly all plot in the low Na/O group. This implies that Type III and low Na/O materials are intrinsic to the long-period comet ejecta and are not the product of more recent heating as meteoroids in the interplanetary medium.

All sodium poor or sodium free meteors have low beginning heights (height class type III) and the meteoroid itself, or a significant fraction of its grains, seem to have experienced high enough temperatures to lose sodium-containing minerals at some time in the past. Not all low beginning height meteors are sodium poor/free, perhaps because not all heating that results in sintering and a density increase leads to the loss of sodium.

### 3.4. Differences in meteoroid density

Figure 8 shows the measured median density of meteoroids in a shower as a function of semi-major axis of the orbit. For all showers reported in Jenniskens (2023), filled circles show the high beginning-height component (height class Type I) and gray diamonds show the low beginning-height component (Type III). Points with open squares and circles are the young and episodic showers of Table 1. The properties of young and episodic shower meteoroids are no different of others of the same dynamical group. Meteoroid density does not change significantly with age.

There is a marked difference between Jupiter-family ("short-period" in Fig. 8) and long-period comet showers. Both showers have Type I and Type III meteors, but in Jupiter-family comet showers the Type I have significantly higher density, almost at the level of Type III materials in long-period comet showers. Median densities for long-period comet showers are $\rho \sim 0.63$ g/cm$^3$ for Type I and $\rho \sim 2.0$ g/cm$^3$ for Type III, while median densities for Jupiter-family comets are $\rho \sim 1.5$ g/cm$^3$ for Type I and $\rho \sim 2.6$ g/cm$^3$ for type III, respectively. That same difference between long-period comet and Jupiter family comet meteoroids is shown for larger meteoroids in terms of the PE criterion, showing that Jupiter-family comet meteoroids can withstand higher dynamic pressures (Borovicka et al., 2022).

Figure 8 also shows that there are some exceptions, but they are relatively rare. The Draconids (IAU shower #9) from comet 21P/Giacobini-Zinner and the Andromedids (IAU#18) from comet 3D/Biela are examples of low-density material among Jupiter-family comets. They are known to readily fragment (e.g., Borovicka et al., 2007; Chenna Reddy & Yellaiah, 2015). Some long-period comet showers also have height class Type I meteors that have a higher density of about 1.2 g/cm$^3$. One example being the $\alpha$-Monocerotids (IAU#246), which has $\rho = 1.16$ g/cm3 for Type I.

### 3.5. Size frequency distribution of meteoroids

Turning back to the meteor magnitude distribution index $\chi$ in a shower, we find typically higher values for Jupiter-family comets compared to long-period comets, again with some exceptions. This is shown in Fig. 9, which plots the young (episodic) showers as filled circles, while long-period comet showers for which $\chi$ has been corrected for age to the time of ejection, are given as gray crosses.

Table 2 counts the number of showers in each comet population assuming that $\chi$ is similar to four values expected for different physical processes: $\chi = 1.85$, 2.15, 2.50, and > 3.0 respectively.

Most long-period comet showers eject meteoroids with a low $\chi \sim 1.85$, a value typical of collisionally relaxed populations where each log mass interval has a constant amount of surface area (Jenniskens, 2006). We see that trend in both in young showers and in the population of age-corrected long-period comets (gray crosses in Fig. 9). Examples from young showers include the Filament components in the showers of Halley-type comets 1P/Halley ($\chi \leq 1.85 \pm 0.10$), 55P/Tempel-Tuttle ($\chi$ = 1.50 ± 0.10), 8P/Tuttle ($\chi$ = 1.75 ± 0.05), and 109P/Swift-Tuttle ($\chi$ = 1.77 ± 0.05).

Young Jupiter-family comets, on the other hand tend to have $\chi$ ~ 2.15, typical for a collisional cascade, where growth would be accompanied by gentle collisional fragmentation in a way such that big particles break up into smaller pieces, and then those smaller pieces break and become the parents of even smaller pieces, etc. (Jenniskens, 2006). That, again, is most apparent in the meteoroids trapped in mean-motion resonances, like the well observed s-Taurid shower #628 from 2P/Encke ($\chi$ = 2.11 ± 0.12) (Asher & Izumi, 1998; Devillepoix et al., 2021), the h-Virginids (shower #343, $\chi$ = 2.31 ± 0.03), the $\eta$-Virgnids (#11, $\chi$ = 2.25 ± 0.04), the $\mu$-Virginids (#47, $\chi$ = 2.05 ± 0.10), the June xi$^1$-Sagittariids (#861, $\chi$ = 2.14 ± 0.17), the June Bootids (#170, $\chi$ = 2.33 ± 0.21), and the Camelopardalids (#451, $\chi$ = 2.37 ± 0.17).

In other cases, however, the magnitude distribution index of young Jupiter-family comets is steeper, with $\chi$ ~ 2.52 that implies constant mass per log-mass interval, as expected for an equilibrium population where meteoroids are catastrophically destroyed in collisions, or even $\chi$ > 3.0, indicative of meteoroids that fragment abundantly by processes other than collisions (Fig. 9). Well observed examples are the $\chi$-Cygnids (shower #757), with $\chi$ = 3.11 ± 0.05, and the Andromedids (#18) with $\chi$ = 3.33 ± 0.10.

There are again some exceptions. The kappa-Cygnids (shower #12) are rich in bright fireballs with $\chi$ = 1.74 ± 0.02, but are on a short-period Jupiter-family comet type orbit (Fig. 9). These fireballs also stand out from other such showers in having multiple flares, indicative of relatively weakly consolidated material (Jenniskens & Vaubaillon, 2008; Jenniskens, 2023).

Nearly all JFC showers with low meteoroid densities have large $\chi$ > 3.0. Figure 10 compares the magnitude distribution index with the bulk density of meteoroids in a meteor shower, restricted to comet showers with q > 0.3 AU. Jupiter-family comets are shown as open symbols, the long-period comet showers as solid dark symbols. The showers with $\chi$ > 3.0 are the showers that have the lower meteoroid densities.

### 3.6. The grain mass distribution in individual meteoroids.

The meteor light curve shape parameter (F) is thought to reflect the mass distribution index of individual grains that make up the meteoroid ($\chi_g$), with lower F corresponding to higher $\chi_g$. The grains are not necessarily distributed the same as the mass distribution index of the population of the meteoroids in a stream ($\chi$). The latter reflects the final state of pebbles just before the collapse to comets, while the grain mass distribution index reflects the process of particle growth leading up to that moment. We like to explore whether or not they are correlated.

Figure 11 shows this relationship for height class type I meteors with $q > 0.3$ AU. JFC showers are marked as "o" and LPC showers as "•". Age-corrected LPC showers are shown as a gray cross. The solid particle value of $F = 0.76$ (classical light curve) is shown as a vertical dashed line. Both groups are limited to showers for which the F-parameter and the magnitude distribution index were both derived from the young component only, where that component is well enough isolated from the annual background.

Among most long-period comets (•, in Fig. 11), there is no correlation between light curve shape parameter F and the magnitude distribution index. In this case, the meteoroid building blocks are not distributed in the same way as the meteoroid population as a whole.

Among Jupiter-family comets (o, in Fig. 11), higher-$\chi$ showers have lower F values, as expected if there is a correlation between $\chi$ and $\chi_g$, albeit that the values do not match to those calculated by Beech & Murray (2003) for $\chi = \chi_g$ to hold, perhaps only because our meteoroids are larger than those considered by Beech & Murray.

Some of the age-corrected LPC showers (gray crosses), those with high $\chi$, scatter among the JFC showers. Those high $\chi$ showers tend to be the older showers (see Fig. 9, where they tend to have low median semi-major axis). We interpret this to mean that change of $\chi$ from aging (Jenniskens et al., 2004) does not change the $\chi_g$ of the meteoroids.

### 3.7. Admixture fraction of type III materials ($f_{Ib}$)

The young episodic component of meteor showers is typically less well observed than the older shower components that return annually, which makes it more difficult to measure the admixture fraction of type III materials. For that reason, we show here both the young showers and all other detected showers with perihelion distance $q > 0.3$ AU from which at least $N > 100$ meteors were detected. However, the admixture fraction of Type III materials can be affected by age if Type I materials are lost from the stream faster than Type III materials.

Figure 12A shows the admixture fraction as a function of the semi-major axis (a) of the orbit. Streams with a < 5 AU are short period showers, most JFC. Filled and open circles are young showers. The age-corrected showers are shown as a gray cross.

Figure 12B shows that fraction as a function of the inclination of the orbit. Now, open points are the JFC-type showers with semi-major axis a < 5 AU, while gray "x" symbols mark all JFC type showers. Solid points in Fig. 12B are the young LPC type showers, while gray crosses show all LPC type showers.

Young LPC mostly have admixture fractions of Type III materials around 4%, distributed in a relatively narrow range (increasing to slightly higher values for older showers with smaller semi-major axis). That range is larger than the formal uncertainty (1 sigma), with some solutions significantly above the majority, such as the $\alpha$-Monocerotids (shower #246, at the top of Fig. 12B), which has a high admixture fraction of Type III materials.

Jupiter-family comets have on average higher admixture fractions of ~8%, but with a much wider range of values than found for LPC showers (Fig. 12B). Very few Jupiter-family comet showers have admixture fractions as high as the 50–65% admixture fraction derived from Stardust samples of comet 81P/Wild 2 (Westphal et al., 2009).

There is also a population with low 0.3–0.8% admixture fractions for both groups of comets, both in low and high inclinations, also seen among several young showers (Fig. 12B). We checked to see if that could be due to the limit of N > 100 set. Weak showers do not contribute to detections below the minimum admixture fraction value of 1% (ratio is 0.01). However, there are not fewer long-period comet showers with a large enough number of observed meteors (N > 1000).

Showers in prograde high inclined orbits (Mellish-type showers – Jenniskens, 2023) contain higher ~13% admixture fractions. Those orbits tend to show large Kozai cycles with perihelion distance approaching the Sun in one part of the cycle, suggesting an effect of aging, with Type I materials being lost from the stream faster than the more compact Type III materials (Jenniskens et al., 2024).

The long-period comet population as a whole (gray crosses) shows a weak trend of showers with a lower median semi-major axis orbit having a higher admixture fraction (Fig. 12A). This, too, is due to aging. A lower median semi-major axis orbit points to a higher age of the stream. The parent LPC have semi-major axis in the range 40 to 250 AU (orbital periods 250 to 4,000 y,

Jenniskens et al., 2021). The young LPC showers (solid circles in Fig. 12A) mostly have these higher median semi-major axis.

### 3.8. Primitive asteroids

Only four of the young meteoroid streams in Table 1 are from primitive asteroids (all with short semi-major axis) and all were likely produced by disruptions close to the Sun. All have a high magnitude distribution index $\chi \sim 2.94 \pm 0.22$, and an admixture of Type III materials more typical of Jupiter-family comets than long-period comets. However, heating by the Sun can have changed these properties. In this case, the meteoroid densities are not low (Table 1) and the high magnitude distribution index is likely intrinsic to the debris created during disruption.

## 4. Discussion

### 4.1 Changes from comet internal radiogenic as well as geochemical heating

Our premise is that that meteoroids still contain information about their accretion conditions. Recently, Malamud et al. (2022) have argued that internal radiogenic as well as geochemical heating will mobilize volatiles and so there are no pristine comets, but they also found that aqueous alteration that could affect the internal cohesion, density and composition of individual pebbles (e.g., Rubin et al., 2007) is not expected for comets smaller than about 20 km in diameter. Most of our meteor showers ae caused by the more common smaller comets. Perseid comet 109P/Swift-Tuttle, with a 26-km diameter (Jenniskens, 2006), may be the exception but shows no evidence of aqueous alteration or high cohesion in its meteoroids. Hence, it is possible that the thermal evolution of comets in the few million years after formation did not remove the signature of the original pebble pile.

### 4.2. Changes during and after ejection

*4.2.1 The meteoroid density.*

The most striking difference between the observed LPC and JFC meteoroids is the higher meteoroid density for JFC showers compared to those from LPC and HTC (Sect. 3.4). This was first noticed by Kikwaya et al. (2011) from a small sample of meteor light curve modellings, albeit that their study had average densities of $1.1 \pm 0.8$ g/cm$^3$ for HTC type meteoroids and $3.1 \pm 0.3$

g/cm$^3$ for JFC meteoroids, with no overlap between the populations. We find here that the densities are lower and that there are exceptions in both populations (Sect. 3.4). Our densities are scaled to molten silicates being the highest density values, and agree well with early values reported by Buccongello et al. (2023) from high spatial resolution meteoroid fragmentation modeling. The specific showers studied in Buccongello et al. (2024) have nearly the same densities to those calculated here for the height class Type I shower meteors (Jenniskens et al., 2024). However, the average bulk densities for each comet group reported by Buccongello et al. (2024) are a factor of two lower than reported here: 0.345 ± 0.048 g/cm$^3$ for HTC showers and 0.602 ± 0.155 g/cm$^3$ for JFC showers. Both average values are in the same ratio, so it is possible that our density values are systematically overestimated.

Does the density difference between LPC and JFC relate to an initial different fraction of ices in the material? The observed meteoroids have lost their ices (and perhaps some volatile organics) so they are not the same material anymore as the pebbles that collapsed to form the comet, but perhaps not too different. That is because the fraction of ice to solids in comets is low (see Sect. 4.2 below). If lower density is on account of having lost a higher fraction of volatiles (ices), then that would imply LPC contain a significantly higher fraction of ices compared to JFC. To explain a meteoroid density difference of 0.63 g/cm$^3$ for LPC and $\rho \sim 1.5$ g/cm$^3$ for JFC (Sect. 3.4) from the fractional loss of water ice, that would imply that LPC contain about 74% ice and JFC only 38%, if both contain rocky material of the grain density 2.42 g/cm$^3$ of carbonaceous chondrite meteorites (Macke et al., 2010). That, too, points to different source regions for both populations.

Does the density difference reflect the number of times that the comet has approached the Sun? Jupiter family comets are dynamically older than Oort cloud comets. The typical Jupiter-family comet that produces meteor showers on Earth has completed hundreds of orbits in the inner solar system, whereas Oort-cloud comets are known to be affected by rapid fading and experienced relatively few perihelia passages. Perhaps not. Meteoroids are thought to be ejected from near the surface of the comet because they have sufficiently high internal cohesive strength and are already in a suitable size for gas drag, but with sufficiently low tensile strength between meteoroids (Gundlach et al., 2015). Sintering from solar radiation increases the cohesion among meteoroids and prevents ejection, creating insulating surface layers.

Alternatively, could the density difference be the result of space weathering of the meteoroid after ejection while in orbit around the Sun? This does not seem to be the case (Sect. 3.4), with young and old showers showing similar densities for each height class (Fig. 8).

*4.2.2 The magnitude distribution index.*

Also different between JFC and LPC showers is the magnitude distribution index $\chi$ (Sect. 3.5). Could this be an effect of conditions during ejection of the meteoroids? Gundlach et al. (2020) modeled the ejection of meteoroids from Jupiter-family comet 67P/Churyumov-Gerasimenko and concluded that the sublimation of $CO_2$ ice drives the ejection of large ($\geq$ 10 cm) chunks (dominating the mass loss), still containing 10-90% of water ice, while the sublimation of $H_2O$ ice causes the lift-off of pebble-sized meteoroids, which contain no ice. However, the chunks will fall apart later when the ice sublimates, leaving the same size distribution for all ejecta.

Jupiter-family comets tend to be less active than LPC, but it is unclear how that would affect the $\chi$ of the ejected meteoroids. A weaker gas drag would be expected to cause less fragmentation, rather than more. Moreover, a higher density of JFC meteoroids would argue for less violent fragmentation and, again, a shallower mass distribution, rather than a steeper one.

If the weaker gas drag is not able to lift the larger grains, and if the material would start with the same distribution of pebbles and more of the smaller pebbles were ejected as meteoroids, then that would leave the larger pebbles on the comet, which over time would create a thick crust that is not observed.

*4.2.3 The admixture fraction of type III and low sodium materials.*

In addition to the density and $\chi$, other differences between the LPC and JFC populations encountered here include the slightly higher diversity of admixtures with sodium-poor and relatively dense materials in the comet matter for JFC, with LPC showing a narrow range and more consistent values (Sect. 3.3). With JFC being dynamically older, could they accumulate more material at the surface and modify this material before ejection to add Type III meteoroids to the stream (assuming modification at the comet surface would be different from modification while in space as a meteoroid)? In that case, being close to the Sun might be a stronger contributor than the number of times that the comet was in the inner solar system (Jenniskens et al., 2024). We examined if there was a correlation between the admixture fraction and the perihelion distance for the range $0.3 < q < 1.02$ AU, but found none for both young JFC, young LPC, or old LPC.

### 4.3 Changes from comet orbit and meteoroid stream dynamical evolution

Given that the Sun's radiation can alter the grain morphology if the perihelion distance $q < 0.3$ AU (Jenniskens et al., 2024), how well does the current perihelion distance describe the perihelion distance of past orbits of the meteoroids? The young JFC and LPC showers did not significantly change q since ejection, but could orbital dynamics account for sodium-poor meteoroids in some of the older LPC showers?

Most LPC streams never evolved below $q < 0.3$ AU, even though they likely evolved for a period of ~50,000 y before being detected at Earth. In Fig. 13, we show a number of integrations of LPC configured to match known showers having a selection of inclinations and perihelion distances. We created comet bodies with the same elements as the showers, and adjusted the eccentricities to produce initial periods either of 4000 years or 250 years. Integration was performed forwards from the present time, using an ephemeris following (Fienga et al., 2008), and osculating orbital elements output versus time. The elements of the showers discussed below are tabulated in Jenniskens (2023), and our numerical algorithms are described in (Jenniskens et al., 2024). In all cases with $q > 0.3$ AU, the perihelion distance is seen to remain relatively constant over a 50 kyr integration. At long periods, the comets interact with the inner solar system only infrequently. When a comet does interact, it is traveling fast enough that its orbital energy cannot reduce significantly, and so the period remains long.

Only if the orbital period of the parent body decreased to below about 100 years do we see mean-motion resonances come into play that can change the parent body perihelion distance. With such short orbital period, the comet interacts many times with the inner solar system, and has low enough energy that it can be pulled into a short orbit and fall prey to resonances. These are the Halley-type comets (HTC). They are relatively rare, but they can produce active showers. For that reason, all known HTC showers have been studied well and none had a past evolution to $q < 0.3$ AU since ejection (e.g., Jenniskens, 2006).

### 4.4. Corroborating evidence for differences between LPC and JFC

*4.4.1 The nucleus density of comets and the dust-to-gas ratio.*

Is there supporting evidence for the observed difference in the bulk density of LPC and JFC meteoroids seen here? Meteoroids near the surface of Jupiter-family comet 67P/Churyumov-Gerasimenko have a typical density of 0.8 g/cm$^3$, but with a wide range of 0.001 – 4 g/cm$^3$ (Jorda et al., 2016). No such measurements exist for long-period comets.

Another approach is to compare the bulk density of cometary nuclei, assuming the macro-porosity is the same. Table 3 is a literature search of such information, building on Blum et al., 2006. Comet 67P has a bulk density of 0.532 ± 0.007 g/cm$^3$ (Jorda et al., 2016), only slightly lower than the typical density of 0.8 g/cm$^3$. There is much uncertainty about the refractory-to-ice ratio in 67P, with most estimates a high ≥ 3 (Choukroun et al., 2020), so that the meteoroid densities after evaporation of the ice are expected to be similar to those of the bulk comet density (in the absence of macro-porosity). 67P is not the parent of one of our meteor showers.

There are no estimated nucleus densities of long-period comets, but comet 1P/Halley is estimated to have a nucleus bulk density of 0.47 ± 0.19 g/cm$^3$ (Rickman, 1989). For the Orionids (shower #8), we have a meteoroid density of 0.41 g/cm$^3$ for the fluffy Type I meteoroids and a fraction of 9.5% of compact Type III materials with $\rho$ = 1.60 g/cm$^3$, for a mean density of 0.52 g/cm$^3$. The dust-to-ice ratio in this comet is highly uncertain, but likely >> 1, making it, again, not unreasonable that the meteoroid density is similar to the nucleus density as a whole.

Other data seems to confirm that the bulk densities of the two populations are similar, but JFCs slightly higher. Recently, Paradowski (2022) determined using indirect methods the nucleus density for 136 comets and found a mean bulk density of 0.47 ± 0.01 g/cm$^3$ for 40 long-period comets, lower than the 0.54 ± 0.01 g/cm$^3$ for 94 short-period comets.

If the two comet populations have nearly the same bulk densities, then our results imply that Jupiter-family comets contain more macro-porosity (on a scale >> 1 cm) than long-period comets.

*4.4.2 Contact binaries.*

It was shown that a clump of pebbles in the streaming instability would shed its angular momentum by forming an equal-mass equal-color binary, which can collapse to form a contact binary (Nesvorny et al., 2010). Table 3 suggests that another possible difference between JFC and LPC is that not all JFC are contact binaries. Contact binary 19P/Borrelly has a low nucleus density of 0.24 ± 0.06 g/cm$^3$ (Davidsson & Gutiérrez, 2004). Jupiter-family comets that are not contact binaries include 9P/Tempel 1: 0.60 ± 0.40 g/cm$^3$ (Davidsson et al., 2007; Richardson et al., 2007), and 81P/Wild 2: 0.70 ± 0.10 g/cm$^3$ (Davidsson & Gutiérrez, 2006), which both have slightly larger nucleus density. All values are lower than the densities measured here for the JFC meteoroids. None of these particular comets is the parent body of one of our meteor showers.

*4.4.3 The D/H ratio.*

Other evidence for differences in the two comet populations comes from the isotope ratio of deuterium over hydrogen (D/H) of water ice, half the measured value of the HDO/$H_2O$ ratio (e.g., Altwegg et al., 2015; Anderson et al., 2022). Although there may be effects from the nature of the comet activity (e.g., Fulle, 2021), that ratio is higher for LPC on average than for known JFCs, with some exceptions (Hallis, 2017; and Table 3). Long-period comets 1P/Halley, C/1996 B2 (Hyakutake), C/1995 O1 (Hale-Bopp), 2002 T7 (LINEAR), 2001 Q4 (NEAT), and 8P/Tuttle all have D/H ~ 3 x $10^{-4}$ within uncertainty intervals, while JFC comets 103P/Hartley 2 and 45P/Honda-Mrkos-Pajdušáková have D/H ~ 1.6 x $10^{-4}$, similar to Earth's oceans (Vienna Standard Mean Ocean Water: D/H ~ 1.56 x $10^{-4}$). The water in CI and CM ordinary chondrite meteorites has a slightly lower D/H (Peslier et al., 2017). Again, exceptions exist. Jupiter-family comet 67P/Churyumov-Gerasimenko has a high D/H = 5.01 ± 0.10 x $10^{-4}$ (Müller et al., 2022).

## 5. Implications for the conditions in the solar nebula at the time of pebble collapse

Can we understand the differences between LPC and JFC if they originate from their formation conditions? Comets were formed in the protoplanetary disk through some leapfrog growth mechanism such as the Streaming Instability (Youdin & Goodman, 2005), or through the enhanced concentration of pebbles in turbulent eddies (known more formerly as Turbulent Concentration, Cuzzi et al., 2008; Hartlep & Cuzzi, 2020) and collapse into comets and asteroids. Our hypothesis is that the meteor shower observations show the state of the solid material at the time of this collapse.

### 5.1 The mass distribution index

Our observations show that most LPC formed from material that accreted in gentle accretional processes, meaning, in the parlance of coagulation models, that they formed in regions of the disk in which growing particles did not reach the fragmentation barrier but were possibly subject to bouncing. The process of growth is determined by the surface area of the particles. This resulted in a magnitude distribution index of $\chi \sim 1.85$ (Table 1, Fig. 9), which corresponds to a size distribution index $\alpha = 3.0$ (and differential mass distribution index of $s = 1.67$), with most mass in the large particles. The shallow slope for the LPCs (Sect. 3.5) may indicate a more characteristic slope for fragmentation free particle growth in this part of the proto-planetary disk. We also find that there is no correlation between meteoroid size distribution index and the size distribution of the grains that make up the meteoroid. Those grains tend to have steeper size distributions (lower F). Comets 1P/Halley and 81P/Wild 2 show similar steeper size distributions for smaller < 1-mm meteoroids (Hörtz et al., 2006).

We also find that, in contrast, most Jupiter-family comets formed in regions of the proto-planetary nebula where the particle growth was limited by fragmentation at the time of primary accretion. The higher $\chi \sim 2.15$ (Sect. 3.5) corresponds to $s = 1.83$ and size distribution index $\alpha = 3.5$. This suggests a more violent collision environment due to higher particle-to-particle relative velocities (e.g., Pan & Padoan, 2013) or other forces capable of fragmenting the particles, where particles were approaching or may have even reached the fragmentation barrier. If particles did reach the fragmentation barrier, then this would have almost certainly affected the slope of the particle size distribution and may account for some showers with $\chi \sim 2.52$ among Jupiter-family comet showers, corresponding to $s = 2.00$ and size distribution index $\alpha = 4.00$ (Table 2). Whether pebbles are in a regime of fragmentation or not depends on ambient nebula conditions, but generally models show that grain growth is less likely to reach the fragmentation barrier the further out in the protoplanetary disk one is, unless the disk is very turbulent (e.g., Estrada et al., 2022).

### 5.2 The density of the meteoroids

The difference in density between JFC and LPC shower meteoroids (Sect. 3.4) could have come from different growth processes during pebble formation. The pebble density, before the collapse into JFC, could have increased through compaction in collisions. The compaction could have been stronger if relative velocities where higher, or the ice content was lower. These conditions could reflect the normal collisions as the pebbles grew larger, or the gas density decreased. The JFC formation region could have had higher turbulent velocities or lost more water from higher particle or pebble temperatures compared to the LPC formation region. This would account for both fluffy and dense materials having increased their density in Jupiter-family comet matter.

Additional compaction might have occurred during the collapsing cloud of pebbles, where collisions within the cloud were dominated by bouncing, but collision velocities might not have been much greater than those from the local turbulence before collapse. Once these clumps form and collapse, the relative velocities generally do not exceed the escape velocity of the cloud (Lorek et al., 2016).

Finally, compaction of pebbles may have occurred after the formation of comets. Every single object could have experienced stochastic collisional evolution at different stages in its early evolution (e.g., Blum et al., 2006; Morbidelli & Nesvorny, 2020). If so, the higher density of the meteoroids would suggest more collisions among Jupiter-family comets than among long-period comets.

**5.3 Low beginning height and low sodium meteoroids**

We find that both Jupiter-family comets and long-period comets contain Type III materials with low beginning height, at least some of which are poor (or free) in sodium (Sect. 3.3). The loss of sodium requires that these dense materials were thermally altered by warming. We see the loss of sodium for streams with perihelion distances $q < 0.2$–$0.3$ AU (Jenniskens et al., 2024), where meteoroids are heated to about 500–600 K, but also find such materials in streams on orbits that never evolved below $q = 0.3$ AU, and concluded that these materials are intrinsic to the comet. For reference, Mercury's orbit has a perihelion at $q = 0.308$ AU. Indeed, the light scattering properties of comet matter requires both fluffy and dense materials in the comet coma and in-situ missions show both fluffy and dense materials on the comet surface (e.g., Levaseur-Regourd et al., 2018).

Given that low-Na materials are found in meteoroids on a cm-size scale, and those meteoroids have slightly higher densities and start to ablate deeper into the Earth's atmosphere, the initial pebbles were likely heated when they had grown to this size. In analogy with loss of sodium in our Solar System today, that would suggest that these pebbles were heated over extended periods of time to above 600 K.

Melting alone is not sufficient. Chondrules are often not depleted in sodium (Hewins, 1991). Ablation of Na in micrometeorites during the rapid meteor phase starts only at ~1700 K, while Fe ablation starts at ~1800 K with some minerals ablating not until ~2100 K (Bones et al., 2019). These values correlate with equilibrium condensation temperatures for a Solar System composition gas (depending on $O_2$ fugacity), ~958 K for Na, ~1336 K for Mg, and ~1334 K for Fe (Lodders, 2003). However, to make meteoroids sodium poor does not require the sublimation of sodium-containing minerals, but rather could mean the loss of small grains rich in Na. Trigo-Rodriguez et al. (2004), Trigo-Rodriguez & Llorca (2007) and Trigo-Rodriguez & Blum (2022) pointed out that sodium is sometimes overabundant in meteoroids and may been mixed with volatile organic matter in the protoplanetary nebula. Some evidence for that comes from Stardust samples of comet 81P/Wild 2 (e.g., Matrajt et al., 2008) and possible cometary inclusions in meteorites (Nittler et al., 2019). Organic matter pyrolysis in small meteoroids was measured to start earlier at around 625 K (Bones et al, 2022), and it is possible that Na is lost partially from fluffy meteoroids when organic matter is lost at this temperature and the meteoroid lost some of its glue.

The different admixture fractions of ~8% for JFC and ~4% for LPC also point to different source regions. It is possible that low-Na materials originated from near the Sun. Turbulence and diffusion in the models of protoplanetary disks would predict a certain level of mixing of mm- and cm-sized pebbles. Transport from the inner solar system would likely result in smaller pebbles being added to the population further from the Sun, and the gradient would likely be steep. Stellar jets could also carry inner-solar pebbles outward. In both scenarios, the abundance of compact Na-poor materials is a function of heliocentric distance. Our observations tend to measure slightly larger sized meteoroids for JFC showers on account of their lower entry velocity in Earth's atmosphere, and their near constant numbers with LPC showers could hint at larger pebbles having been supplied to the JFC origin region.

It is perhaps more likely that moderately volatile element were lost from pebbles by disk winds from the proto Sun at high altitudes from the disk plane (Sengupta et al., 2022). That, too, would create less Na-poor pebbles at high heliocentric distances.

The wide range in admixture fractions found in JFC could mean that these comets formed over a range of heliocentric distances, or over a more extended period of time than LPC. The relatively narrow range of admixture fraction in LPC could mean that these comets were formed during a relatively short period of time.

## 6. Implications for the origin region of LPC and JFC

In past studies, the LPC were thought to have originated in the giant planet formation region, while JFC were thought to have originated from the Trans Neptunian Disk. In modern dynamical models, that picture has changed. Now, dynamical models argue that both LPC and JFC derived from the Trans Neptunian Disk. The Kuiper Belt's Scattered Disk, which is the reservoir of Jupiter-family comets stretches up to the Oort cloud and feeds it (Fernandez et al., 2024; Nesvorny, 2018). That would predict no significant differences between JFC and LPC. Our observations, however, show that there are intrinsic differences between JFC and LPC.

### 6.1. Origin of Long-Period Comets in the proto-solar nebula

Here, we propose that the differences between JFC and LPC are because the current Oort cloud population (source of LPC) survived only from late stages of solar system evolution compared to the Scattered Disk population (source of JFC). That possible Big Picture is schematically summarized in Figure 14, adapted from Morbidelli & Nesvorny (2020), depicting in Fig. 14A the conditions in the protoplanetary disk at the time of the photoevaporation of the gas and formation of planetesimals, in Fig. 14B the migration of planets due to accretion of planetesimals until the point of a dynamic instability that increased the inclination and eccentricities of the giant planets, and in Fig. 14C the post-instability regime and final outcome of planetesimal accretion. The Classical Kuiper Belt is thought to be a remnant of the proto-planetary disk, which extended beyond the 30 AU edge of the Trans Neptunian Disk that stopped Neptune from moving further out.

The Oort Cloud is thought to have formed when planetesimals had strong interactions with the young gas giants and galactic tides were able to lift their perihelion distance out of the inner solar system (Oort, 1950; Shoemaker & Wolfe, 1984). Passages of nearby stars and galactic tides then randomized the angular orbital element distributions. However, when that happened is still debated. Some alternatives have been proposed that differently identify which giant planets scattered the comets and when. Encounters with the young Uranus and Neptune are the most efficient at putting objects in near-parabolic orbits (Fernandez, 1980). It is possible to have more planetesimals captured from the Jupiter/Saturn region of the proto-planetary disk if forces in the Sun's stellar birth cluster are considered (e.g., review of Gladman & Volk, 2021). Our observations seem to argue against that, as most Oort Cloud comets appear to have formed under conditions where the growth of meteoroids that assembled them was not fragmentation limited.

The Oort Cloud comets can also have originated from the exchange of materials between the Sun and the planetary disks of its sibling stars as they formed and drifted apart (Eggers et al., 1997; Levison et al., 2010). However, that would perhaps have resulted in more diversity in the admixture of compact materials in the cometary streams. Hence, our observations also argue against this scenario.

Instead, our findings are consistent with a scenario where the Oort Cloud formed well after the Sun moved away from its birth star cluster. Early on, the gas drag would have prevented ejection into Oort-Cloud orbits, and as long as the Sun was still in the star cluster the gravity of nearby stars would have put comets in unbound orbits (Brasser et al., 2006; Morbidelli & Nesvorny, 2020). The Oort Cloud could have formed at the time of the dynamical instability that increased the eccentricity and inclination of the giant planets to its current values, long after the gas was removed (Thommes et al., 2002; Tsiganis et al., 2005). This event now is thought to have occurred about 60–100 Ma after the beginning of the solar system (Morbidelli et al., 2023) and was perhaps associated with the ejection of a planet from the system as a result of planet migration (Fig. 14C). At the time, Neptune had evolved to about 27.7 AU, assuming the 1:2 mean-motion resonance cleared the area up to 42 AU as indicated by a kernel in the Kuiper Belt Object (KBO) distribution (Petit et al., 2011).

The mechanism that creates the Scattered Disk and the Oort Cloud is the same: perturbations by Neptune. In this scenario, the Oort Cloud is formed mostly from the planetesimals that remained in the 27.7 to 30 AU part of the solar nebula and some objects in the disk beyond (Fig. 14). Following the instability, the planet migration included at least a period of gradual outward migration of Neptune from 27.7 to 30 AU that did not disturb too much the classical Kuiper Belt at ~42–47 AU (e.g., review by Gladman & Volk, 2021). However, it is not clear what happened to the 30 to 42 AU region that is now void of Classical Kuiper Belt Objects.

This far from the Sun, it is not unreasonable to find that long-period comets mostly were created under gentle accretion conditions and with similar admixture fractions of compact materials. The higher D/H ratio of most long-period comets compared to JFC must mean that their pebbles formed in conditions where the gas density and temperatures were lower, because there was less isotope exchange between gas (low in D) and solids (high in D). The gas density would have fallen off towards the outer edge of the Trans Neptunian disk, with lower gas density to higher heliocentric distances. It is tempting to think that the planetesimals beyond the 30 AU edge of the disk, between 30 and 42 AU, were the most well located to keeping the high D/H ratio in solids, but it is unclear if sweeping resonances from Neptune during the instability would later have brought enough planetesimals to Neptune to populate the Oort Cloud and not disturb too much the classical Kuiper Belt.

If the long-period comets come from this region, then that could also explain the discontinuity between the properties of Jupiter-family and long-period comets. In this scenario, perhaps the best analog to the long-period comets would be the objects that are currently still in the Classical Kuiper Belt. With classical KBO object Arrokoth having a bulk density of ~ 0.24 g/cm$^3$, with 1 sigma range 0.16–0.6 g/cm$^3$ (Keane et al., 2022), the meteoroids generated from such an object after evaporation of the ices would be similarly low if the ice to solid ratio is low.

### 6.2. Origin of Jupiter-Family Comets in the proto-solar nebula

Planetesimals that populated the Scattered Disk originated from the whole primordial Trans Neptunian Disk, along Neptune's entire migration path. This could naturally explain higher densities and a wider range of admixture fractions of compact (low-Na) materials.

It is unclear at what distance from the Sun the planet Neptune initially formed, but it has been assumed at ~15–25 AU (Morbidelli & Nesvorny, 2020; Eriksson et al., 2023). Planet migration from accretion and scattering of planetesimals caused Neptune to continue to move out and large KBO's like Pluto to be trapped in mean-motion resonances with Neptune and be included in the Scattered Disk population (Malhotra, 1993; Dones et al., 2015; Hadden & Tremaine, 2024). The Kuiper Belt's dwarf planets are found in the Scattered Disk. These objects have different reflection properties and contain less similar-mass and similar-color binaries than those of the Classical Kuiper Belt because of close encounters with Neptune (Morbidelli & Nesvorny, 2020). It is thought that the current Scattered Disk is the 1% that remains today of the population of objects scattered by Neptune since the beginning of the Solar System and which have not found a stable parking orbit (Duncan & Levison, 1997; Morbidelli & Nesvorny, 2020).

At the time of the planet upheaval and subsequent late migration when the Oort Cloud was formed, some planetesimals were scattered also into the Scattered Disk. Hence, it is no surprise that low-density and high D/H interlopers are found in the Jupiter-family comet population, including comets 67P/Churyumov-Gerasimenko and 21P/ Giacobini-Zinner.

The prevalence of contact binaries among JFC and KBO, as well as LPC (Table 3), supports the idea that streaming instabilities or turbulent concentration (or some combination thereof, Sengupta & Umurhan, 2023) in the protoplanetary nebula created by the interaction of gas and solids first brought the cm-sized pebbles together (e.g., Morbidelli & Nesvorny, 2020). Because most JFC originated in the Trans Neptunian Disk closer to the Sun than LPC, it is possible that a more sudden collapse of these instabilities could have resulted in more macro-porosity for JFCs than for LPC.

There is some evidence from modeling that planetesimals in the Trans Neptunian Disk experienced collisions, while the Classical Kuiper Belt population did not (Nesvorny et al., 2020). Indeed, comet 67P appears to have lost a large chunk, possibly in a collision. Also, both JFC and LPC are often contact binaries, but some JFC are not, perhaps having lost one of the components in collisions (Table 3). This argues against the Trans Neptunian Disk population having been in collisional equilibrium at the 1-km size scale, but rare collisions do explain why some JFC would not be contact binaries.

### 6.3. Origin of primitive meteorites in the proto-solar nebula

Planet migration as a result of planetesimal accretion and scattering also occurred early, e.g., the "Grand Tack" migration when Jupiter moved inwards in the first few million years while Uranus and Neptune moved outwards, and thus clearing the asteroid belt in the process, but Jupiter later reversed course and re-populated the asteroid belt with planetesimals from the inner and outer solar system (Walsh et al., 2011). Primitive asteroids were born big (Morbidelli et al., 2009; Ribero de Sousa et al., 2024), but perhaps not bigger than other objects in the Trans Neptunian Disk.

There is a discontinuity in the isotopic compositions of meteorites (e.g., Cr) between enstatite and ordinary chondrites on the one hand (inner solar system) and carbonaceous chondrites on the other (outer solar system), which suggests that the primitive asteroids responsible for the carbonaceous chondrites came from some distance away from the terrestrial planet formation region.

That still leaves a wide range in heliocentric distances as a possible source region for primitive asteroids. Our observations argue against asteroids having been implanted from a disturbed Scattered Disk population during the dynamic instability, or from planetesimals near 30 AU following the dynamic instability (Nesvorny et al., 2008), which would make primitive asteroids the same as JFC. For that same reason, the observations also argue against primitive asteroids having originated from just outside the giant planet formation region early in the Trans Neptunian Disk evolution (Levison et al., 2009).

Guilera et al. (2014) pointed out that aggressive fragmentation of planetesimals (and pebbles) to the point that most mass ends up at the small-size end ($\chi > 2.52$), which they set at 0.01 cm, will benefit the growth of planetary cores that can form the giant planets rapidly.

The one shower from a primitive asteroid with q > 0.3 AU in our dataset (Sect. 3.8) crumbled with a steep mass distribution index. Note that these meteoroids were likely too fragile to be represented in our meteorite collections. If this property is intrinsic to the material in the parent body, then perhaps primitive asteroids formed in the region of giant planet formation (Fig. 14).

Meteorites also provide some insight. The matrix materials and chondrules in CV type carbonaceous chondrite Allende have a similar size distribution when normalized to their modal value (Simon et al., 2018), which could reflect the initial pebble distribution, even if the pebbles were compact aggregates of chondrules. The large-size end of that size distribution corresponds to $\chi \sim 3.17 \pm 0.15$, in agreement with that derived here from our four meteoroid streams of primitive asteroids. If we take this number at face value, then most mass was in the small pebbles when this planetesimal formed.

We also examined the observed ejecta from primitive asteroid Bennu (CI-type carbonaceous chondrite meteorites), which were in the size range of approximately 0.5 to 5 cm, with smaller meteoroids too faint to be observed. Using published volume-equivalent spherical diameters (Chesley et al., 2019), we calculated the expected meteor peak magnitudes for each size range. After examining the fully sampled bright-end of the distribution and accounting for the formal uncertainty in each magnitude bin, we fit an exponential curve to the data. This analysis yielded a magnitude distribution index of $\chi = 2.17 \pm 0.20$ over the range of +1 to -4 magnitude, a value more typical of Jupiter-family comets. Note that this material contained few chondrules and may have originated at the far end of the possible source region (Fig. 14). It is important to note, however, that the pebbles that formed Bennu have undergone significant aqueous alteration since their formation, making it unlikely that this distribution still accurately reflects the initial pebble distribution.

If chondrules are equivalent to the dense materials in cometary matter, then their admixture fraction is perhaps even more variable than in JFC, with chondrules constituting up to 80% of the volume of most primitive meteorites, but CI chondrites (Zande, 2004) and the material of Bennu having few chondrules. Primitive asteroids tend to have higher densities (1.57–3.04 g/cm$^3$) than JFC (Macke et al., 2010). The water in CV, CI and CM chondrites has lower D/H than SMOW, perhaps slightly lower than in JFC (e.g., Peslier et al. (2017). All these arguments support the hypothesis that primitive asteroids formed in the giant planet formation region and not in the Trans Neptunian Disk (Fig. 14).

*6.3.1. Some Oort cloud comets from the giant planet formation region?*

The exceptional α-Monocerotids (shower #246 in Figs. 12) have a very high ratio of Type III materials that are also sodium-poor (Jenniskens et al., 1997). The shower has no known parent body. This adds to the observation that some compact materials on a tens of cm size scale are found in long-period comet orbits (Vida et al., 2023).

There is a category of long-period comets that have an asteroidal (tailless) appearance, called the *Manx* comets after tailless cats of that name, which release ~0.65 g/cm$^3$ meteoroids poor in ice (e.g., Sekiguchi et al., 2018; Kwon et al., 2022). Manx comets would only produce meteoroid streams during breakup events. It is possible that the α-Monocerotids originated from this type of comet.

Unlike the majority of LPC, these comets may have originated from the giant planet formation region. There are several possible pathways how that could happen. One example we mentioned earlier is the possibility to have planetesimals captured into an Oort Cloud from the Jupiter/Saturn region of the proto-planetary disk if forces in the Sun's stellar birth cluster are considered (e.g., review of Gladman & Volk, 2021). Long-period comets like the one that produced the α-Monocerotids might be a remnant of the primordial Oort Cloud that formed when the Sun was still in its birth cloud.

## 7. Conclusions

The meteoroids that cause our meteor showers have properties that relate to the properties of the pebbles that collapsed to form comets and primitive asteroids. The observed meteoroids have a size similar to that of pebbles in observed protoplanetary disks (~1 cm). They have lost their most volatile components, but have densities similar to the comet nucleus density, which suggest pebbles were compact and not fractals and comets are rich in rocky and carbonaceous solids, not ices. The meteoroids do not have a characteristic mass or size, but rather an exponential mass distribution that peaks at either high or low masses outside the range at which meteors are counted during meteor showers. There might be an upper-size cut-off (around 10 cm), from which few meteors are detected. Pebble distributions were likely the same. There were both low-density chondritic abundance pebbles and a small fraction of pebbles of higher density (some of these Na-poor or Na-free), perhaps equivalent to matrix material and chondrules in meteorites.

There are differences between long-period and Jupiter-family comet meteoroids that appear to be due to conditions in the solar nebula at the time of the jump from pebbles to planetesimals. Most long-period comets now release cm-sized pebbles with magnitude distribution index $\chi \sim 1.85$, typical of collisionally relaxed matter with constant surface area per log mass interval. They appear to have formed from pebbles that accumulated under conditions where they were not fragmenting. This squares with the general low density of the material. Their meteoroid mass distribution index does not correlate with their grain mass distribution index, suggesting that their growth conditions varied over the lifetime of the pebble and at the final growth stages pebbles came together gently. They have a nearly constant ~4% admixture fraction of compact and sometimes sodium-poor materials.

Jupiter-family comets, on the other hand, release cm-sized meteoroids with a steeper magnitude distribution index of $\chi \sim 2.15$, or even $\chi \sim 2.5$, suggesting that the pebbles from which they were formed are the product of an environment in which growing particles were close to ($\chi \sim 2.15$) or reached ($\chi \sim 2.5$) their fragmentation barrier. Their meteoroid mass distribution index correlates with their grain mass distribution index, suggesting growth and fragmentation were going hand in hand. The meteoroids contain higher ~8% admixtures of compact and sometimes sodium-poor materials on average, and that admixture fraction varies more widely.

Primitive asteroids (based on only a few streams that formed in proximity to the Sun) release cm-sized pebbles with most mass at small sizes ($\chi >> 2.5$). Meteorites from primitive asteroids have densities higher than measured for JFC meteoroids. Their admixture fraction of chondrules varies widely from containing almost no chondrules (CI type) up to ~80% (CV type).

These observations are consistent with a scenario wherein long period comets formed from pebbles that accumulated under growth conditions which were fragmentation free near the edge of the Trans Neptunian Disk at 27.7–30 AU from the Sun, or in the 30–42 AU region beyond that edge, and were scattered into the Oort Cloud at a late time, when the Sun had left its birth stellar cluster, during or after an instability occurred in the planetary system.

In contrast, Jupiter-family comets were accumulated together with large Kuiper Belt Objects into the Scattered Disk during the full period of Neptune's migration outwards from its birth region.

Most primitive asteroids did not scatter from the edge of the Trans Neptunian Disk, but likely evolved earlier from proto-planets that formed on orbits among the giant planets, where aggressive fragmentation put most mass in the small pebbles.

Each observed population has interlopers that are understood from the dynamics involved. Examples are the Draconids from comet 21P/Giacobini-Zinner, a JFC, that appear to have a density more typical of LPC, and the $\alpha$-Monocerotids from an unknown LPC that appear to have high density and high content of sodium-poor materials more typical of some primitive asteroids.

**Acknowledgements**.

This manuscript benefitted greatly from outstanding reviews by two anonymous reviewers and a review by J. Trigo-Rodriguez. We thank the CAMS station operators on both northern and southern hemispheres for the gathering of data presented here. We thank Jim Albers and Diana Gilbert for the continued operation of the CAMS Spectrograph at the CAMS Sunnyvale station. We thank Matt Tiscareno of the SETI Institute for the initial setup of the PINTEM software, and the SETI Institute for IT support.

**Funding**

This work is supported by NASA's Emerging Worlds grant NNH18ZDA001N. Support for the development and operation of CAMS over the years was provided by earlier NASA grants and individual networks are supported by local grants, notably CAMS Turkey by TUBITAK (Project code MFAG/113F035). The expansion on the southern hemisphere was supported by NASA under Contract NNM10AA11C issued through the New Frontiers Program.

**Data Availability.**

The meteoroid stream orbital elements and dispersions, their magnitude distribution index, light curve shape parameters, densities and fraction of dense materials were published in Jenniskens (2023) and will be made available in digital form through the NASA/PDS node for Small Solar System Bodies.

**CRediT authorship contribution statement**

**Peter Jenniskens:** Conceptualization, Funding Acquisition, Methodology, Data Curation, Formal Analysis, Investigation, Writing – Original Draft Preparation. **Paul R. Estrada:** Conceptualization, Methodology, Writing – Original Draft Preparation, Review & Editing. **Stuart Pilorz:** Investigation, Dynamical Modeling Software, Writing – Original Draft Preparation, Review & Editing. **Peter S. Gural to Steve Rau:** Methodology, CAMS Software, Data Curation, Writing – Review & Editing. In alphabetical order: **Tim Abbott to James D. Wray:** Data Collect & Curation, Writing – Review & Editing. **Dante Lauretta:** Funding Acquisition, Writing – Review & Editing.

**References**.

Table 1. Properties of young meteor showers: The perihelion distance (q), age since ejection, magnitude distribution index ($\chi$), meteoroid density ($\rho$), light curve shape parameter (F), and the ratio of compact Type III over all materials ($f_{lb}$).

| IAU # | Name | Type | q (AU) | Age (y) | $\chi$ Meteoroid | $\rho$ (g/cm$^3$) Meteoroid | F Grains | $f_{lb}$ Meteoroid | Ref. |
|---|---|---|---|---|---|---|---|---|---|
| 13 | Leonids* | LPC/HT | 0.984 | ~700 | 1.50 ± 0.10 | -.- | 0.49 ± 0.15 | -.- | [1] |
| 102 | α-Centaurids | LPC | 0.966 | -.- | 1.68 ± 0.18 | -.- | -.- | -.- | [2] |
| 281 | Oct. Camelopardalids | LPC | 0.991 | ~4,000 | 1.70 ± 0.10 | 0.43 ± 0.11 | 0.50 ± 0.01 | 0.050±0.010 | [3] |
| 923 | 15-Bootids | Mellish | 0.639 | -.- | 1.71 ± 0.18 | ≤1.7 | 0.71 ± 0.06 | < 0.06 | [4] |
| 206 | Aurigids | LPC | 0.677 | ~2,000 | 1.74 ± 0.08 | -.- | 0.57 ± 0.02 | 0.018 ± 0.018 | [5] |
| 15 | Ursids* | LPC/HT | 0.938 | ~1100 | 1.75 ± 0.05 | -.- | 0.59 ± 0.01 | 0.012 ± 0.005 | [6] |
| 535 | θ-Cetids | LPC | 0.514 | -.- | 1.76 ± 0.27 | ~0.90 | 0.66 ± 0.03 | -.- | [7] |
| 7 | Perseids* | LPC/HT | 0.949 | ~3500 | 1.77 ± 0.05 | -.- | -.- | -.- | [8] |
| 246 | α-Monocerotids | LPC | 0.477 | ~1000 | 1.85 ± 0.10 | ~1.16 | 0.73 ± 0.02 | 0.73 ± 0.14 | [9] |
| 8 | Orionids* | LPC/HT | 0.567 | ~3000 | ≤1.85 ± 0.10 | -.- | -.- | -.- | [10] |
| 31 | η-Aquariids* | LPC/HT | 0.595 | ~2400 | ≤2.05 ± 0.09 | -.- | 0.59 ± 0.01 | 0.0027 ± 0.0009 | [11] |
| 107 | δ-Chameleontids | LPC | 0.942 | -.- | 1.87 ± 0.13 | ~0.25 | 0.69 ± 0.03 | -.- | [12] |
| 838 | Oct. δ-Sextantids | LPC | 0.161 | -.- | 1.93 ± 0.30 | ~0.62 | 0.67 ± 0.03 | -.- | [13] |
| 1033 | ω-Carinids | LPC | 0.958 | -.- | 1.98 ± 0.16 | ~0.84 | 0.60 ± 0.02 | -.- | [14] |
| 208 | Sep. ε-Perseids | LPC | 0.708 | ~710 | 1.99 ± 0.21 | -.- | 0.64 ± 0.03 | ≤0.023 | [15] |
| 6 | April Lyrids | LPC | 0.920 | ~2000 | 2.05 ± 0.10 | -.- | 0.55 ± 0.01 | 0.008 ± 0.004 | [16] |
| 184 | July γ-Draconids | Mellish | 0.978 | ~1000 | 2.06 ± 0.17 | -.- | 0.67 ± 0.02 | 0.033 ± 0.033 | [17] |
| 431 | June ι-Pegasids | LPC | 0.899 | -.- | 2.16 ± 0.09 | ~0.67 | 0.62 ± 0.02 | <0.009 | [18] |
| 1034 | γ-Piscis Austrinids | LPC | 0.990 | -.- | 3.16 ± 0.13 | ~0.32 | 0.46 ± 0.13 | 0.057 ± 0.029 | [19] |
| 12 | κ-Cygnids* | JFC | 0.976 | ~3800 | 1.74 ± 0.02 | ~1.08 | 0.66 ± 0.01 | 0.068 ± 0.008 | [20] |
| 137 | π-Puppids | JFC | 0.938 | ≤110 | 1.98 ± 0.11 | -.- | -.- | -.- | [21] |
| 47 | μ-Virginids* | JFC | 1.004 | -.- | 2.05 ± 0.10 | 2.3 ± 0.7 | 0.66 ± 0.01 | 0.139 ± 0.022 | [22] |
| 198 | β-Hydrusids | JFC | 1.014 | -.- | 2.10 ± 0.20 | -.- | -.- | -.- | [23] |
| 628 | s-Taurids* | JFC | 0.372 | ~5200 | 2.11 ± 0.12 | 1.60 ± 0.11 | 0.68 ± 0.01 | 0.028 ± 0.003 | [24] |
| 861 | June $x^1$-Sagittariids | JFC | 0.974 | -.- | 2.14 ± 0.17 | ~1.6 | 0.64 ± 0.03 | 0.045 ± 0.034 | [25] |
| 11 | η-Virginids* | JFC | 0.358 | -.- | 2.25 ± 0.04 | 2.62 ± 0.49 | 0.63 ± 0.01 | 0.003 ± 0.002 | [26] |
| 343 | h-Virginids* | JFC | 0.453 | -.- | 2.31 ± 0.03 | 1.42 ± 0.21 | 0.65 ± 0.01 | 0.052 ± 0.010 | [27] |
| 170 | June Bootids | JFC | 0.340 | ~180 | 2.33 ± 0.21 | ~0.94 | 0.68 ± 0.03 | 0.107 ± 0.065 | [28] |
| 451 | Camelopardalids | JFC | 0.752 | ~120 | 2.37 ± 0.17 | ≤1.8 | 0.54 ± 0.04 | 0.27 ± 0.11 | [29] |
| 108 | β-Tucanids | JFC | 0.978 | -.- | 2.38 ± 0.08 | ~1.8 | 0.66 ± 0.02 | <0.010 | [37] |
| 1046 | 29-Piscids | JFC | 0.966 | ~110 | 2.45 ± 0.38 | ~1.7 | 0.55 ± 0.02 | 0.106 ± 0.034 | [30] |
| 842 | A-Carinids | JFC | 0.977 | -.- | 2.56 ± 0.09 | -.- | 0.51 ± 0.01 | 0.494 ± 0.035 | [31] |
| 1130 | Arids | JFC | 1.001 | ~26 | 2.57 ± 0.08 | 1.8 ± 0.4 | 0.58 ± 0.02 | 0.010 ± 0.010 | [32] |
| 456 | May φ-Scorpiids* | JFC | 0.613 | -.- | 2.60 ± 0.08 | 1.5 ± 0.2 | 0.65 ± 0.01 | 0.140 ± 0.008 | [33] |
| 459 | June ε-Ophiuchids* | JFC | 0.817 | -.- | 2.66 ± 0.16 | ~1.3 | 0.65 ± 0.01 | 0.017 ± 0.010 | [34] |
| 9 | Oct. Draconids | JFC | 0.996 | ~60 | 2.96 ± 0.18 | 0.28 ± 0.04 | 0.51 ± 0.01 | 0.004 ± 0.002 | [35] |
| 758 | Volantids | JFC | 0.973 | -.- | 3.02 ± 0.13 | 2.5 ± 0.7 | 0.64 ± 0.01 | 0.005 ± 0.004 | [36] |
| 1032 | February Hydrids | JFC | 0.811 | -.- | 3.05 ± 0.41 | ~1.3 | 0.61 ± 0.03 | 0.043 ± 0.026 | [38] |
| 757 | χ-Cygnids | JFC | 0.953 | -.- | 3.11 ± 0.05 | 1.34 ± 0.24 | 0.61 ± 0.01 | 0.100 ± 0.008 | [39] |
| 61 | τ-Herculids | JFC | 0.981 | ~27 | 3.21 ± 0.53 | 0.28 ± 0.05 | 0.59 ± 0.01 | 0.011 ± 0.002 | [40] |
| 18 | Andromedids | JFC | 0.801 | ~175 | 3.33 ± 0.10 | 0.33 ± 0.05 | 0.52 ± 0.01 | 0.058 ± 0.006 | [41] |
| 446 | Dec. ϕ-Cassiopeiids | JFC | 0.897 | ~300 | 3.85 ± 0.41 | ≤0.52 | 0.50 ± 0.02 | 0.030 ± 0.015 | [42] |
| 1129 | June $\theta^2$-Sagittariids | JFC? | 0.073 | -.- | 4.68 ± 0.26 | ~0.70 | 0.65 ± 0.04 | 0.908 ± 0.219 | [43] |
| 4 | Geminids | Asteroid | 0.145 | ~1000 | 2.67 ± 0.05 | 1.30 ± 0.03 | 0.63 ± 0.01 | 0.0021 ± 0.0002 | [44] |
| 341 | Jan. χ-Ursae Majorids | Asteroid | 0.221 | -.- | 2.95 ± 0.06 | 3.4 ± 0.3 | 0.55 ± 0.02 | 0.836 ± 0.153 | [45] |
| 924 | 62-Andromedids | Asteroid | 0.395 | -.- | 2.92 ± 0.30 | ≤4.9 | 0.53 ± 0.02 | 0.83 ± 0.14 | [46] |
| 1185 | 87-Virginids | Asteroid | 0.223 | -.- | 3.21 ± 0.17 | 2.1 ± 1.5 | 0.64 ± 0.02 | <0.0078 | [47] |

Notes: *) Filament component of shower (meteoroids in mean-motion resonances);

[1] From 1998 visual observations of Filament, 1994: $\chi = 2.1 \pm 0.3$, 1995: $\chi = 2.0 \pm 0.3$, and 2002: $\chi = 2.0 \pm 0.1$ (Jenniskens, 2006), while Rendtel et al. (2014) have 1998: $\chi = 1.19 \pm 0.02$, 2001: $\chi = 1.60 \pm 0.05$; Single-revolution dust trail encounters show higher values. Light curve shape parameter from Murray et al. (2002);

[2] Recalculated from original visual data in Wood (1980), Jenniskens (1995), and Jenniskens (2006), p. 347–348;

[3] Variable activity of compact annual shower, from all observed CAMS video-detected meteors (Jenniskens, 2023). From video detected meteors during outburst in 2005: $\chi = 1.4 \pm 0.2$ (Jenniskens et al., 2005). Age estimated from it being a one to a few revolution dust trail of a long period comet;

[4] From 18 CAMS-detected meteors of a compact radiant (Jenniskens, 2023);

[5] From 2018 and 2019 outbursts in CAMS video observations when $\chi \sim 3.2$. Fom visually observed 1-revolution dust-trail encounter in 2007: $\chi = 1.74 \pm 0.08$ (Rendtel, 2007a). Other values from visual observations: 1994: $\chi = 1.7$ (Jenniskens, 2006). 2019: $\chi = 2.5 \pm 0.1$ and 2021: $\chi = 1.9 \pm 0.2$ (Jenniskens, 2023);

[6] From CAMS video-detected outbursts in 2017, 2019, 2020, and 2022 around the 2021 perihelion return of 8P/Tuttle. In 2006: $1.9 \pm 0.3$ (Jenniskens et al., 2006). Fireballs detected in 2020 by Pena Asensio et al. (2021). Age estimate from Jenniskens et al. (2007);

[7] Compact core of shower, from CAMS data analysis in Jenniskens (2023), N = 18;

[8] Strong annual background to Filament component of Perseids, also possible contamination from dust trails. From visual observations in 1991: $\chi = 1.96$, in 1992: $\chi = 2.05 \pm 0.05$, in 1993: $\chi = 1.72 \pm 0.02$, in 1994: $\chi = 1.82 \pm 0.05$, in 1999: $\chi < 2.1$, and in 2018: $\chi = 1.75$ (Jenniskens et al., 1998); Rendtel et al. (2014) have 1993: $1.75 \pm 0.05$, 1994: $1.76 \pm 0.02$, and 1997: $1.78 \pm 0.05$;

[9] From CAMS detected 1-revolution outburst in 2019 (Jenniskens, 2023). F value for component Ib;

[10] From periodic broad outbursts on top of, but prior to the peak of, an annual shower component in 1990–1993 and 2005–2008 (Egal et al., 2020a), likely dust trapped in 1:6 and nearby mean-motion resonances with Jupiter. 1993 outburst rich in bright meteors (Miskotte, 1993). Peak of outburst when $\chi$ is at its lowest 2006 (strong peak of ZHR ~ 60/h, on top of annual ~ 15/h): $\chi = 1.58 \pm 0.08$ (Rendtel, 2007b), 2007 (ZHR ~ 70/h): $\chi = 1.85 \pm 0.10$ (Arlt et al., 2008). Age from model by Egal et al. (2020b);

[11] Broad outbursts around solar longitude $\lambda_o = 45^o$ on top of and prior to peak of annual shower component in 1980 and 1993–1997 (Cooper, 1996; Jenniskens, 2006), and from radio forward meteor scatter observations in 2013 and 2017/2018 (online: https://www.iprmo.org/meteor-results/05_eaq/aqr-total-graph.html, last accessed March 8, 2024). Results here are from CAMS-detected meteors in 2013 and 2017/2018 (N = 379), include some annual shower component. Age from model by Egal et al. (2020b);

[12] Compact shower, from CAMS data (N = 16) analysis in Jenniskens (2023);

[13] Compact shower, from CAMS data (N = 12) analysis in Jenniskens (2023);

[14] Compact shower, updated from CAMS data analysis in Jenniskens (2023), now N = 77;

[15] $\chi$ measured here from 2013, 2019 and 2020 video-detected outbursts, but possible contamination from annual component with $\chi = 2.43 \pm 0.03$ (Jenniskens, 2023), while Rendtel et al. (2014) has $\chi = 1.45 \pm 0.14$ from 2013 video observations and $\chi = 2.15 \pm 0.25$ from 2013 visual observations;

[16] Shower known from 1-revolution dust trail encounters with $\chi \sim 2.90 \pm 0.20$ (Jenniskens, 2023). An outburst observed in Japan in 1945 was described as rich in bright meteors (Komaki, 1945). From CAMS derived data (N = 405), activity curve has a secondary peak around $\lambda_o = 32.3^o$ ($32.0^o$ – $32.6^o$) that is rich in bright meteors. Shower is also known for 1-revolution dust trail encounters rich in faint meteors (Porubcan & Stohl, 1981; Jenniskens 2006);

[17] From likely 1-revolution dust trail outbursts in 2016, 2020 and 2021 in CAMS video observations (N = 30). In 2011: $\chi = 1.8 \pm 0.3$ (Holman & Jenniskens, 2012);

[18] Compact shower, from CAMS data analysis in Jenniskens (2023);

[19] Compact shower, annual, from CAMS data analysis in Jenniskens (2023), even more compact outburst in 2021 (N = 15) has same high $\chi = 3.22 \pm 0.37$;

[20] From CAMS detected meteors. Trigo-Rodriguez et al. (2008) estimated origin around 1700, 3800 or 5900 years ago. Jenniskens et al. (2008) has 3600–6000 years ago;

[21] Jenniskens (1995) has from visual observations 1977: $\chi \geq 1.6$, 1982: $\chi \sim 1.9$. Hughes (1992) has 1977: $\chi = 1.98 \pm 0.11$;

[22] Episodic shower, active in most years, evolves in anthelion source and moves out of plane. From CAMS analysis (N = 318) in Jenniskens (2023);

[23] From analysis of visual observations in Jenniskens (2006);

[24] Strong episodic shower. From analysis of CAMS observations in Jenniskens (2023). Devillepoix et al. (2021) has $\chi = 2.0 \pm 0.1$. Age from Egal et al. (2021, 2022 a,b);

[25] 2006 outburst in CMOR (Ye et al., 2016). From Jenniskens (2023) CAMS observations in 2010/2011. 2015, and 2019/2020 (N = 23);

[26] Episodic shower. From Jenniskens (2023) analysis of CAMS observations (N = 617);

[27] Episodic shower. From Jenniskens (2023) analysis of CAMS observations (N = 506);

[28] Occasional outbursts from recent ejecta. From 2022 CAMS observations (N = 28). Rendtel et al. (2014) have 1998: $\chi = 2.22 \pm 0.07$;

[29] Jenniskens (2014; 2023) analysis from 2014 outburst CAMS observations (N = 30). F value for component Ib;

[30] Episodic shower in antapex direction. From 2019 outburst CAMS observations (N = 114) in Jenniskens (2023). From number ratio relative to background sporadics: $\chi = 2.9 \pm 0.3$;

[31] Episodic shower. From CAMS observations in 2014, 2019, 2020, 2023 (N = 587) in Jenniskens (2023). F value for component I;

[32] From 2021 outburst CAMS observations (N = 96). Encounter with several recent dust trails with 1988: $\chi = 2.52 \pm 0.20$, 1995: $\chi = 2.28 \pm 0.11$, 2008: $\chi = 2.37 \pm 0.45$, and 2014: $\chi = 2.70 \pm 0.12$ (Jenniskens, 2023);

[33] Annual shower with some weak variation of activity and radiant arcing up from anthelion source. From CAMS observations in Jenniskens (2023);

[34] Episodic shower. From 2019 CAMS observations (N =177). Jenniskens (2023) has $\chi = 2.46 \pm 0.14$ from all CAMS detected meteors (N = 254);

[35] From CAMS 2018 outburst observations. Literature values summarized in Jenniskens (2023), most plot around $\chi = 3.1$;

[36] From CAMS 2020 and 2021 outburst observations analyzed in Jenniskens (2023);

[37] Episodic shower on top of a background from shower #130. From CAMS observations, improved on Jenniskens (2023) by adding 2024 data and better isolating the background shower (N = 75);

[38] Detected in 2013, 2018 and 2023 (Jenniskens, 2023) (N = 46);

[39] From CAMS detected outbursts in 2015 and 2020, with N = 899 (Jenniskens, 2023). Arcs from anthelion source upwards in radiant latitude;

[40] Periodic outbursts. Data from 2022 CAMS detected outburst from 1995 ejecta (Jenniskens, 2023).

[41] Annual shower with variable activity in CAMS data. Results from Jenniskens (2023), with summary of literature-derived values;

[42] Same parent body as Andromedids. Episodic activity (Jenniskens, 2023);

[43] Episodic shower, high eccentricity orbit, possibly Jupiter-family comet shower (Jenniskens, 2023). F value from IIb component;

[44] Annual shower, young, shows variation of $\chi$ with position of Earth's in its path because unlike other showers we meet this stream at high true anomaly (Jenniskens, 2023). Meteors are type II. Low $\chi$ component after main maximum. F value from IIa component;

[45] Annual shower, compact and strong. F value from IIa component (Jenniskens, 2023);

[46] Annual shower, compact (Jenniskens, 2023);

[47] Annual shower on top of strong anthelion background, compact, all meteors of type II (N = 128). Updated with 2022 and 2023 data from Jenniskens (2023).

Table 2. The number of young meteor showers (N) that show different collision and fragmentation conditions based on their magnitude distribution index ($\chi$). The three populations of primitive asteroids (AST), Jupiter-family comets (JFC) and long-period comets (LPC) are shown separately.

| **Cause of $\chi$** | **$\chi$** | **$N_{AST}$** | **$N_{JFC}$** | **$N_{LPC}$** |
|---|---|---|---|---|
| Collisional relaxed | 1.85 | 0 | 1.5 | 13.5 |
| Collisional cascade | 2.15 | 0 | 5.5 | 4.5 |
| Catastrophic collisions | 2.52 | 0.5 | 9 | 0 |
| Fragmentation by processes other than collisions, loss of large meteoroids | >2.7 | 3.5 | 8 | 1 |

Table 3. Bulk nucleus density, shape, and D/H ratio of comets.

| **Comet** | **Nucleus Density ($g/cm^3$)** | **Type** | **Notes** | **D/H x$10^{-4}$ *** | **Notes** |
|---|---|---|---|---|---|
| **LPC** | | | | | |
| Arrokhot | 0.24 (0.16–0.6) | Contact Binary | [1] | -.- | |
| C/2012 F6 (Lemmon) | -.- | -.- | -.- | 6.5 ± 1.6 | [16] |
| 8P/Tuttle | -.- | Contact Binary | [2] | 4.09 ± 1.45 | [17] |
| C/1995 O1 (Hale-Bopp) | -.- | (Contact) Binary | [3] | 3.30 ± 0.80 | [18] |
| 1P/Halley | 0.47 ± 0.19 | Contact Binary | [4] | 3.1 ± 0.4 | [19] |
| C/1996 B2 (Hyakutake) | 0.3 ± 0.1 | -.- | [5] | 2.9 ± 0.10 | [20] |
| C/2009 P1 (Garradd) | -.- | Contact Binary? | [6] | 2.06 ± 0.22 | [21] |
| C/2014 Q2 (Lovejoy) | -.- | -.- | -.- | 1.4 ± 0.4 | [16] |
| C/1983 H1 (IRAS-A.-A.) | ~0.45 | Contact Binary | [5] | -.- | |
| **JFC** | | | | | |
| 67P/Churyumov-Gerasimenko | 0.532 ± 0.007 | Contact Binary | [7] | 5.0 ± 0.1 | [22] |
| 45P/Honda-Mrkos-Pajdušáková | (~0.19) | Contact Binary | [8] | <2.00 | [23] |
| 46P/Wirtanen | ~0.23 | Contact Binary?/ hyperactive | [9] | 1.61 ± 0.65 | [24] |
| 103P/Hartley 2 | 0.3 ± 0.1 | Contact Binary / hyperactive | [10] | 1.61 ± 0.24 | [25] |
| 81P/Wild 2 | 0.70 ± 0.10 | Single Object | [11] | 1.18–4.14 | [26] |
| 19P/Borelly | 0.18-0.83 | Contact Binary | [12] | -.- | |
| 9P/Tempel 1 | 0.60 ± 0.40 | Single Object | [13] | -.- | |
| 2P/Encke | 0.75 ± 0.25 | Contact Binary? | [14,15] | -.- | |
| 26P/Grigg-Skjellerup | 1.0 ± 0.3 | -.- | [14,15] | -.- | |

Notes: *) D/H water in Vienna Standard Mean Ocean Water (VSMOV) is 1.5576 ± 0.0001 x $10^{-4}$. Sources: [1] Keane et al. (2022); [2] Harmon et al. (2010); [3] Sekanina (1997), Marchis et al. (1999); [4] Rickman (1989); [5] Harmon et al. (1999); [6] Ivanova et al. (2014); [7] Jorda et al. (2016); [8] Sosa & Fernández (2009); Farukhi (2017); DiSanti et al. (2017); [9] Sosa & Fernández (2009); Morton (2018); [10] Thomas et al. (2013); [11] Davidsson & Gutiérrez (2006); [12] Davidsson & Gutiérrez (2004); [13] Davidsson et al. (2007); Richardson et al. (2007); [14] Davidson (2006); [15] Harmon et al. (1999); [16] Biver et al., 2016; [17] Villanueva et al. (2009); [18] Meier (1998); [19] Eberhardt et al. (1995); Martin (1996); [20] Bockelée-Morvan et al. (1998); [21] Bockelée-Morvan et al. (2012); [22] Müller et al. (2022); [23] Lis et al. (2013); [24] Lis et al. (2019); [25] Hartogh et al. (2011); [26] McKeegan et al. (2006).

Figure 1 – Median meteoroid diameter for 0-magnitude meteors in individual meteor showers, taking into account the measured densities. Dashed lines give the size range corresponding to +5 to -5 magnitude over which the magnitude distribution index was measured.

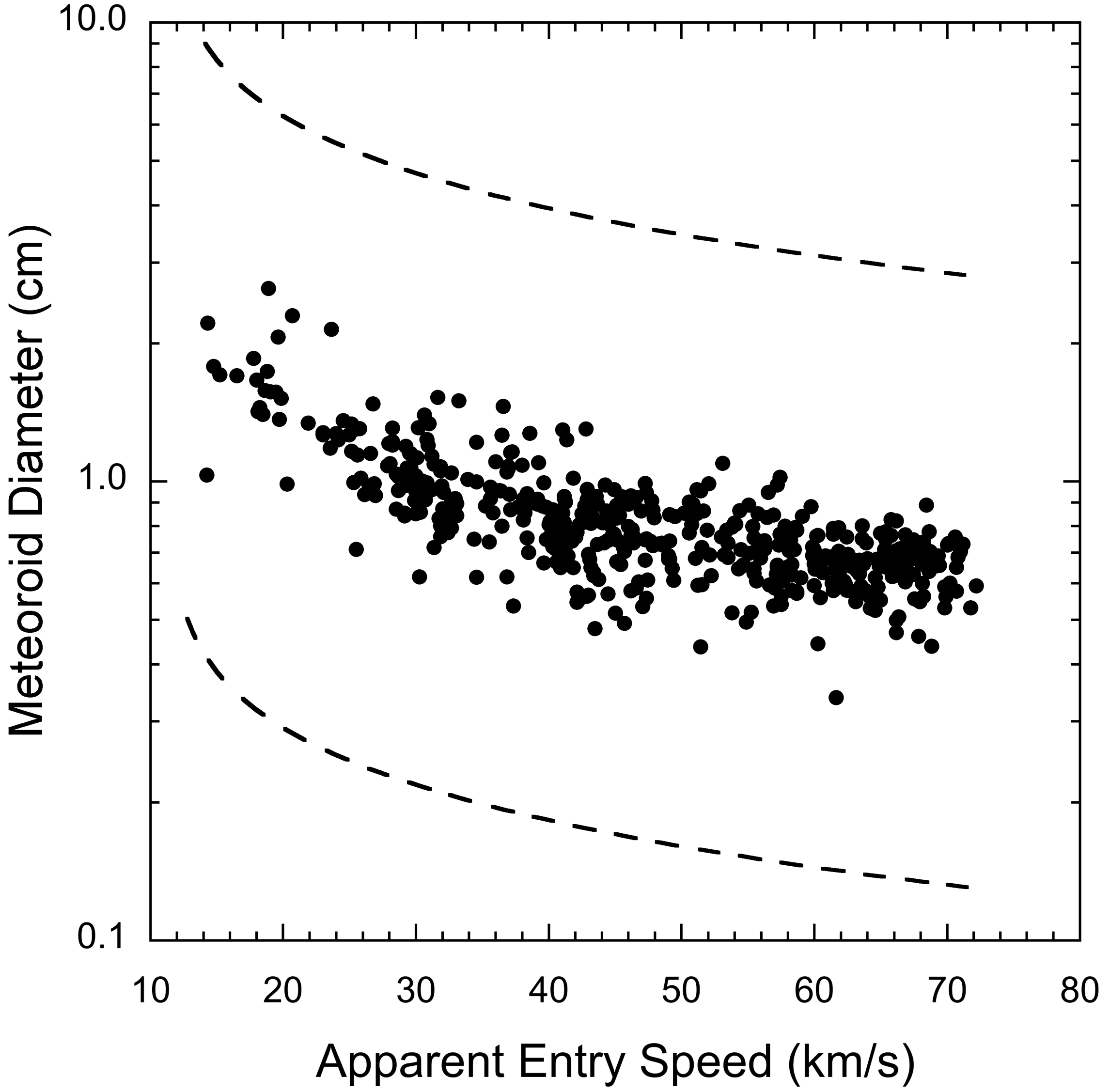

Figure 2 – Examples of corrected magnitude distributions for two observed showers with N = 100 and two cases of slopes χ ~ 2.0 (64 Draconids) and χ ~ 4.1 (December phi-Cassiopeiids).

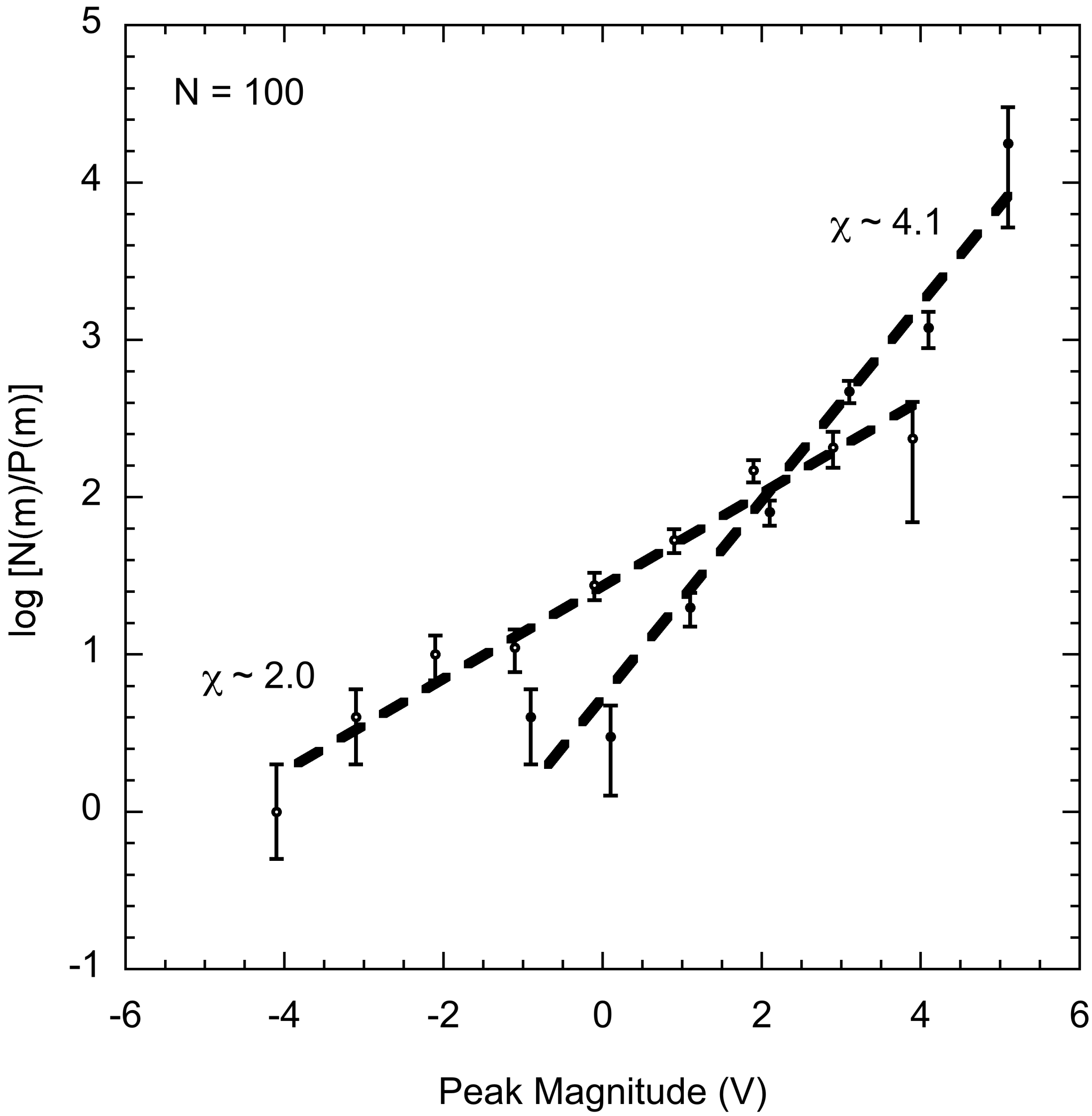

Figure 3 – Example of individual bulk densities measured for meteoroids of the A-Carinids shower (shower #842). Densities in each group (height class Type I and III, separated by a dashed line) do not strongly correlate with height.

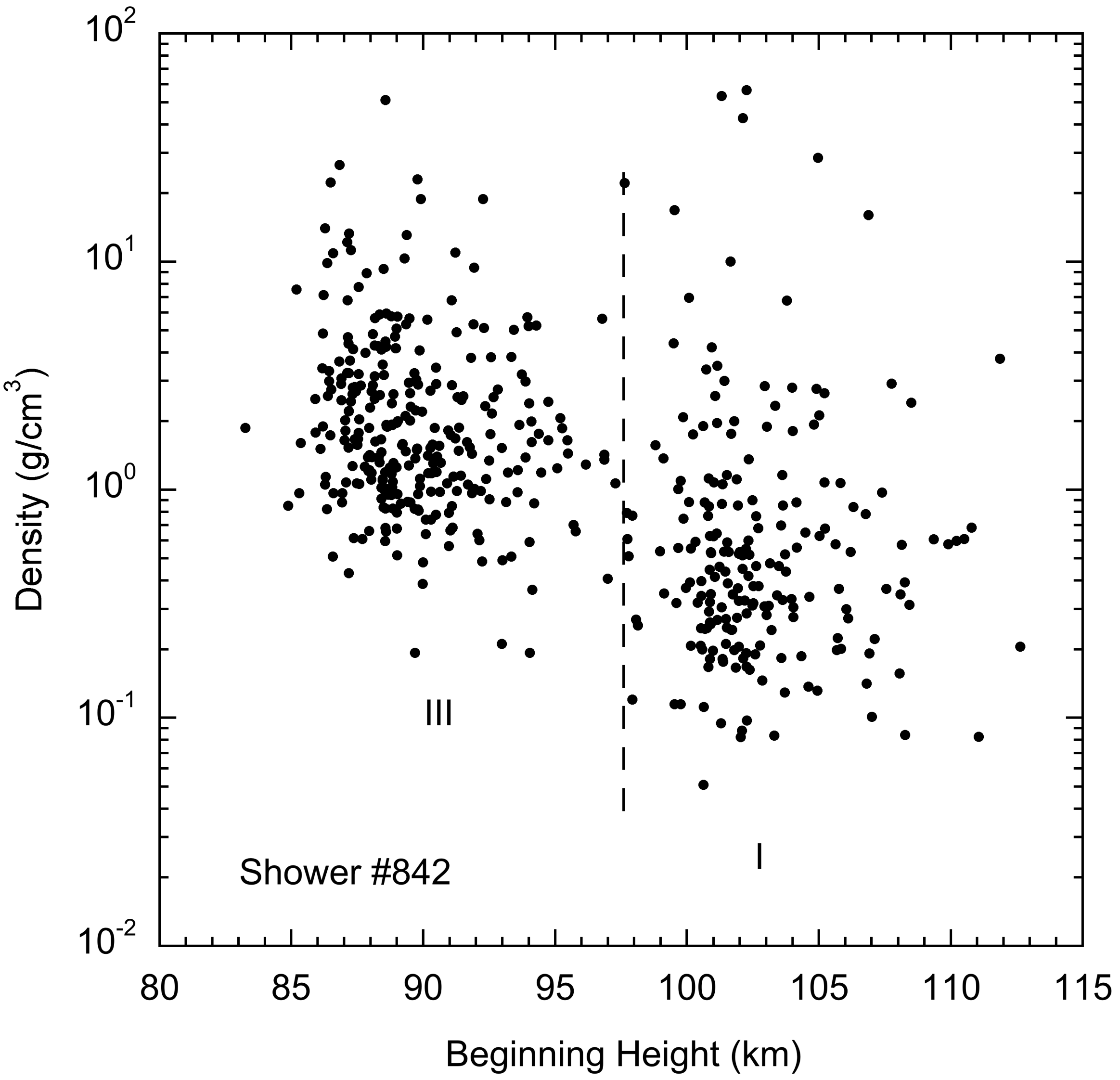

Figure 4A – The CAMS Spectrograph (CAMSS), consisting of 16 spectrographic cameras that operate in conjunction with the CAMS California meteor triangulation network.

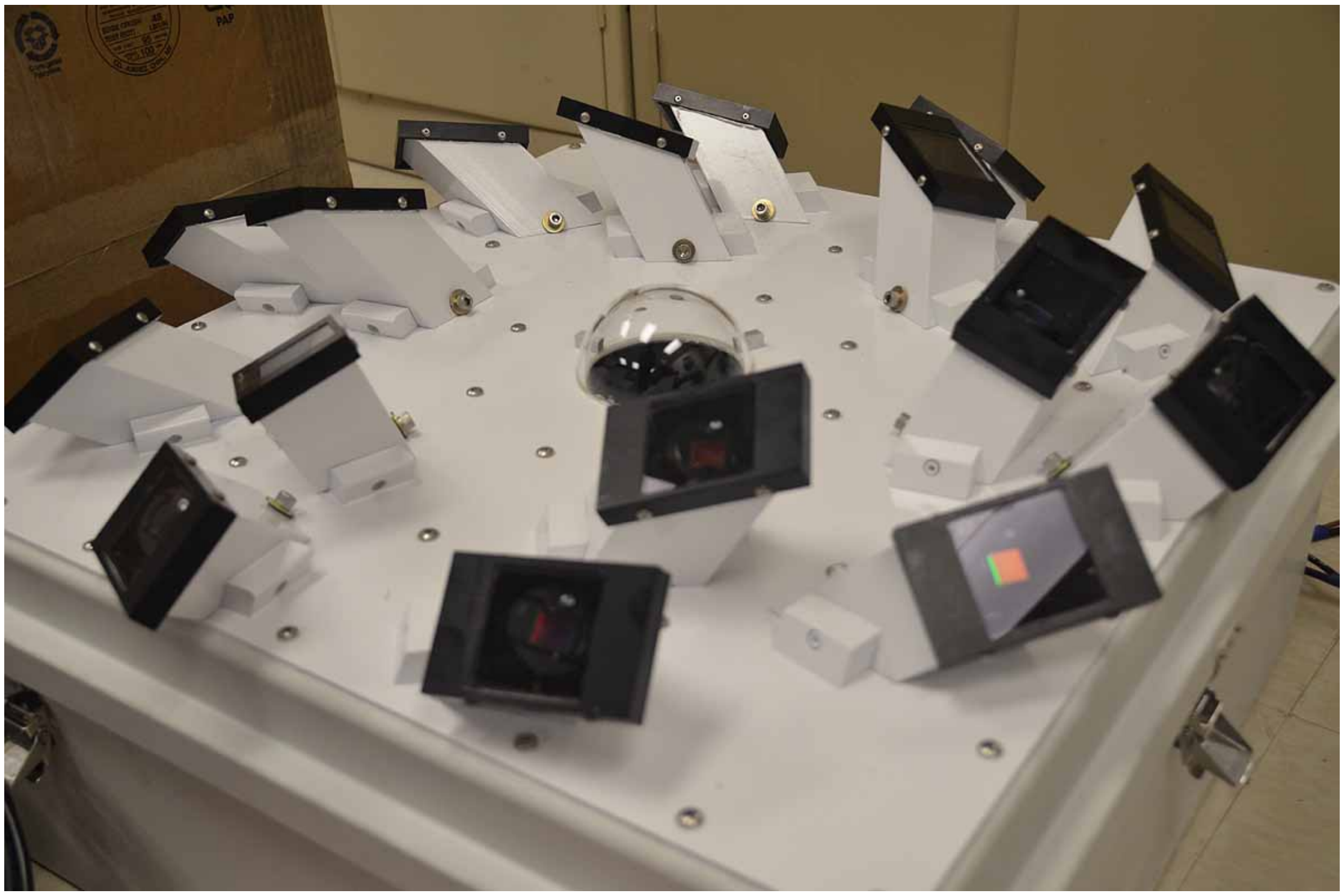

Figure 4B – Screen shot of CAMS Spectrograph reduction software SP_Coincidence, used to extract the meteor spectra from video observations and derive elemental abundances and elemental light curves. Top diagram shows the extracted spectrum (white line), with a wavelength scale that runs from 350 to 660 nm (top) and 640 to 950 nm (bottom) in first order. Emission line fiducials are shown at the top. The blue line is a model spectrum. Bottom left and bottom right diagrams show the parameter adjustment controls for user-selectable parameters, including the grating roll, pitch, yaw, resolution and line width, and the excitation temperature of warm and hot components, the ratio of hot/warm components, the column density, and the altitudes that tie spectrum to CAMS-triangulated trajectory. The middle windows are for element selection to fit individual elements (top) and the CAMSS-CAMS Coincidence info (bottom), while the right window shows a single frame of the meteor spectrum or video.

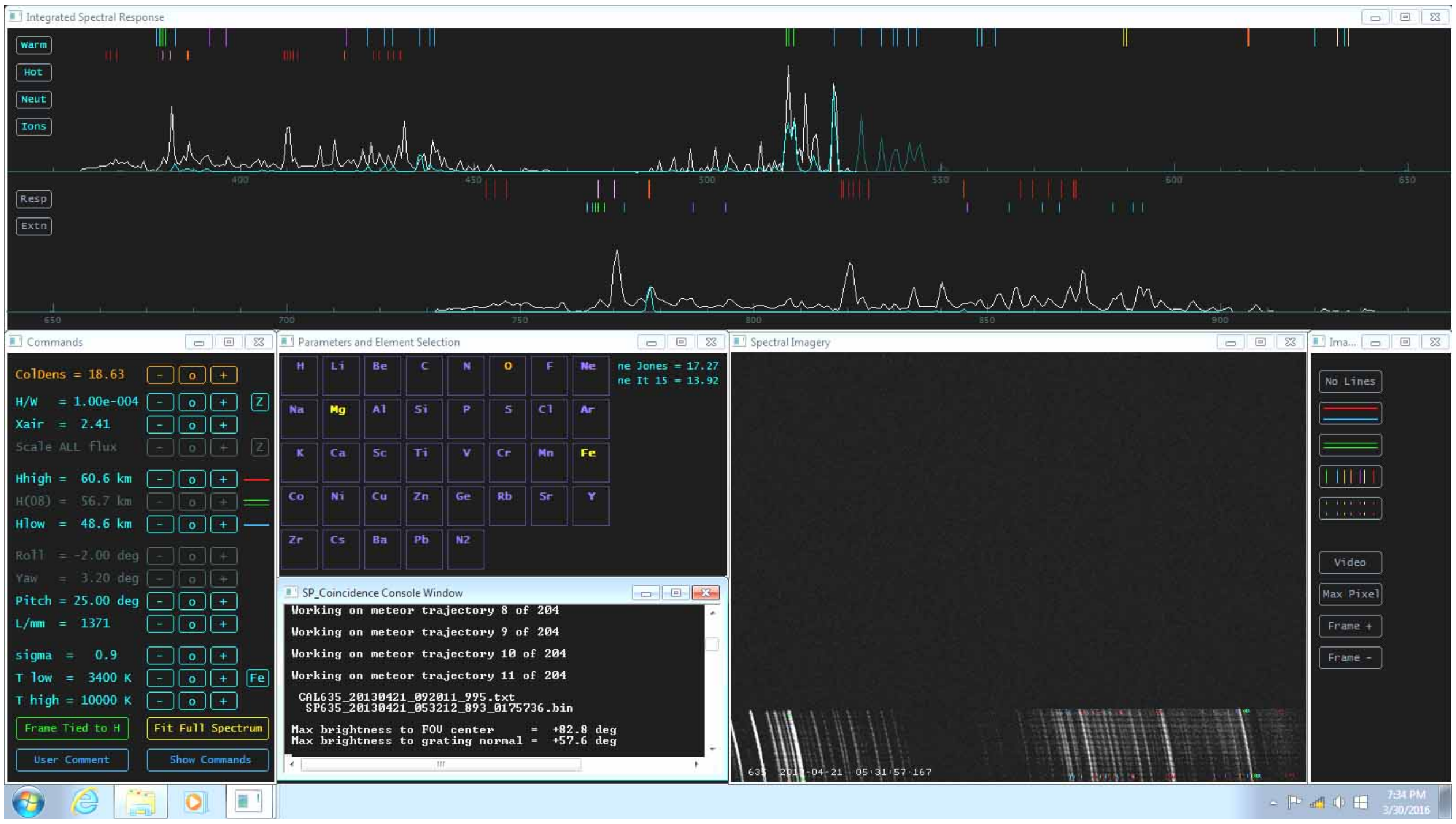

Figure 5A – Line intensities of metal atoms Mg, Fe and Na relative to that of oxygen (after correction for instrument responsivity and telluric extinction) for all spectra for which both one or more of the metal atoms and the oxygen atom lines were measured. Meteors from the active Geminid (GEM) and Perseid (PER) meteor showers are marked.

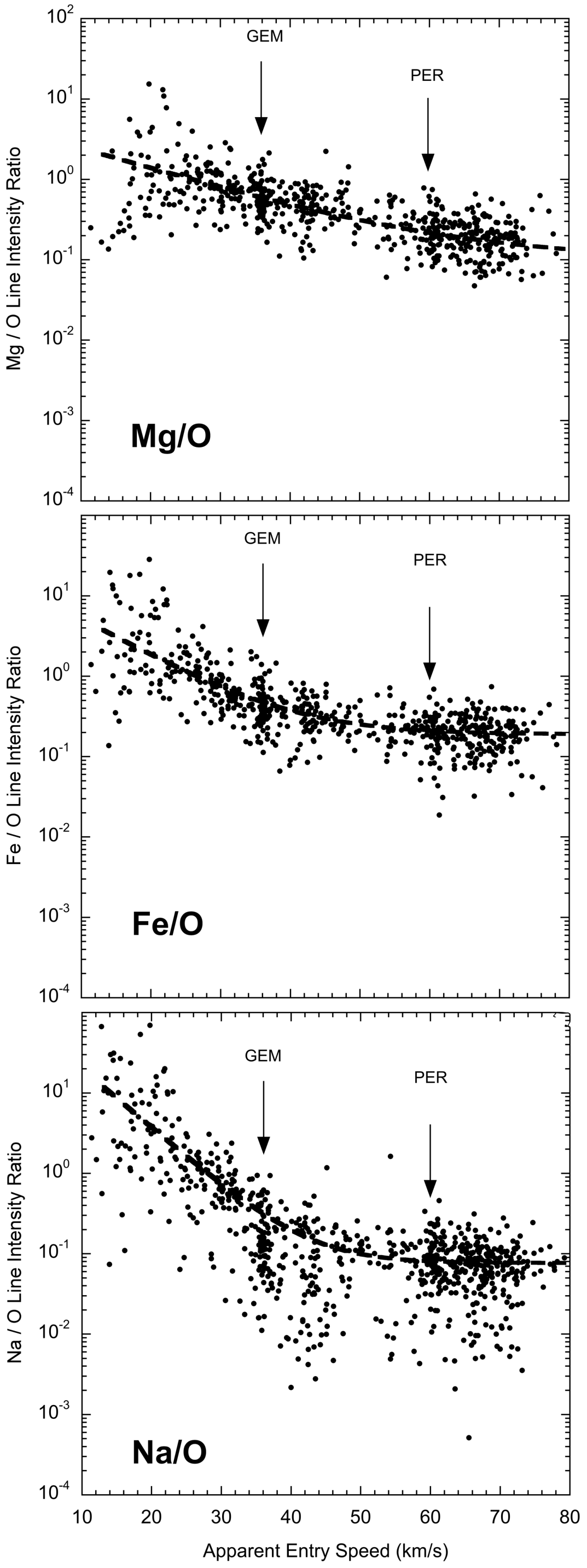

Figure 5B – Line intensities of Fe and Na relative to Mg.

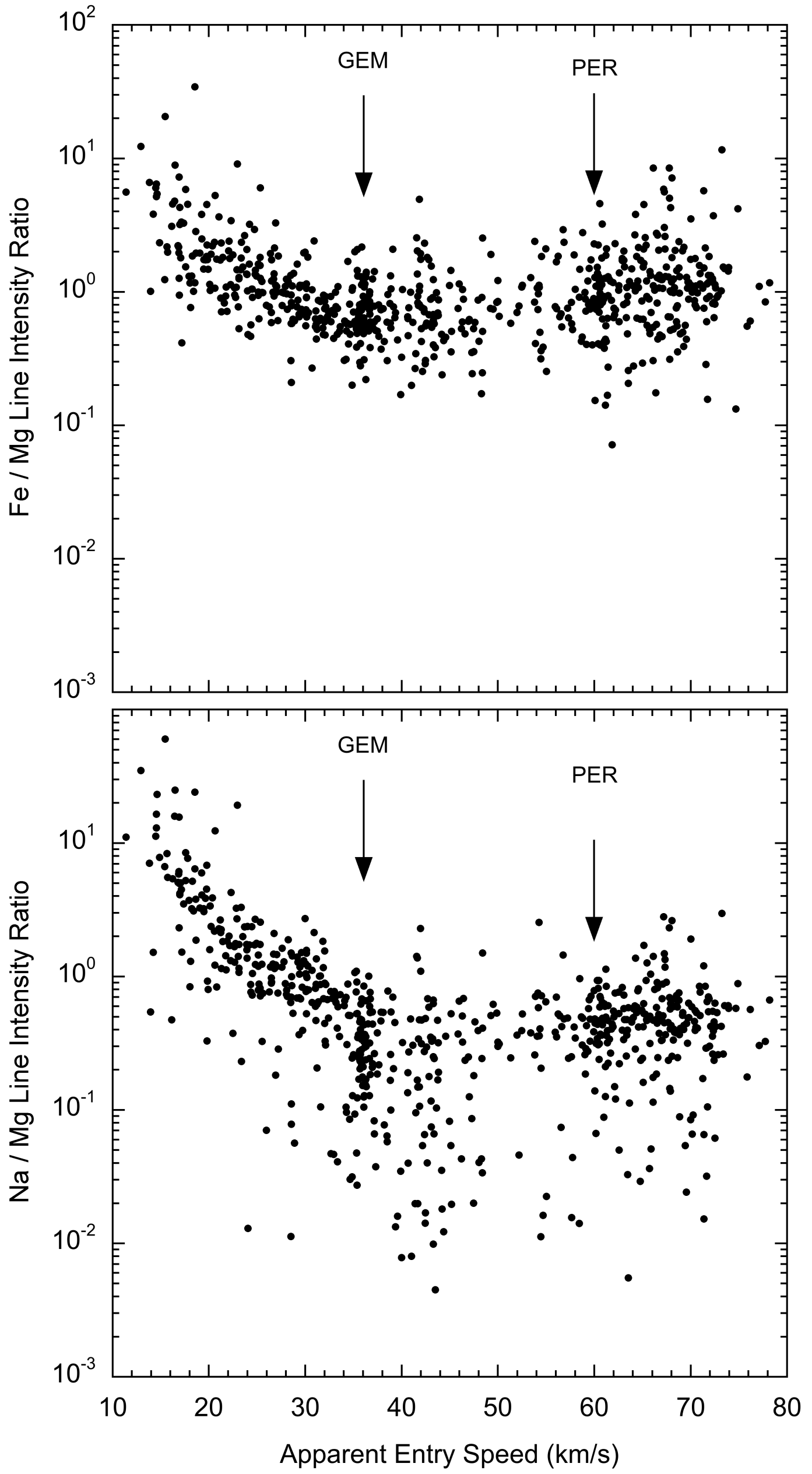

Figure 6 – Beginning height (corrected for velocity-dependence to Height Class parameter $k_c$) as a function of apparent meteor entry velocity for all meteors measured by spectroscopy. Open circles are showers with perihelion distance q < 0.3 AU. Height Classes Type I and III are marked.

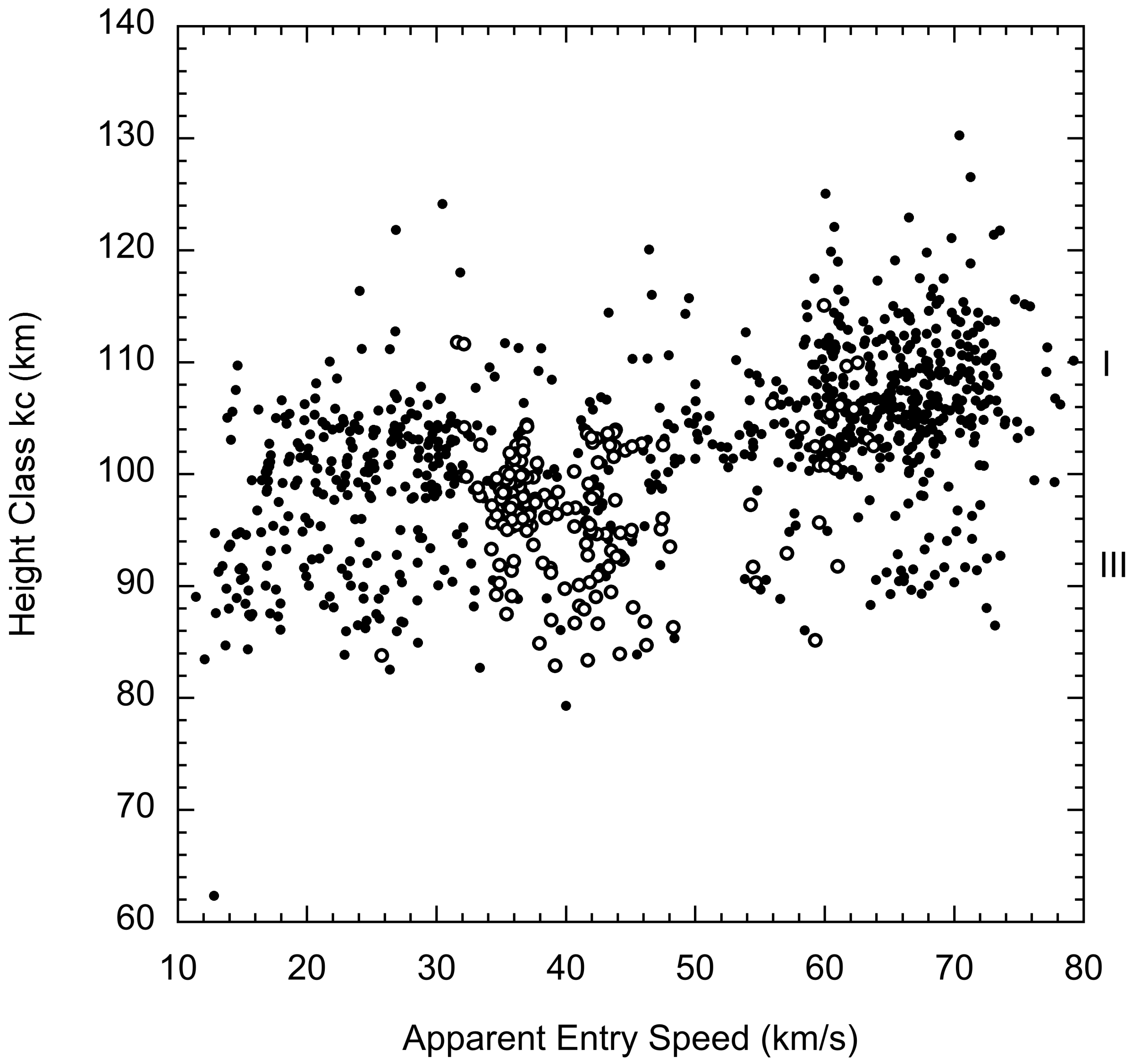

Figure 7A – Relative strength of Na emission versus Height Class parameter $k_c$ for all 1005 spectra measured. Meteoroids with $q < 0.3$ AU are marked as open circles.

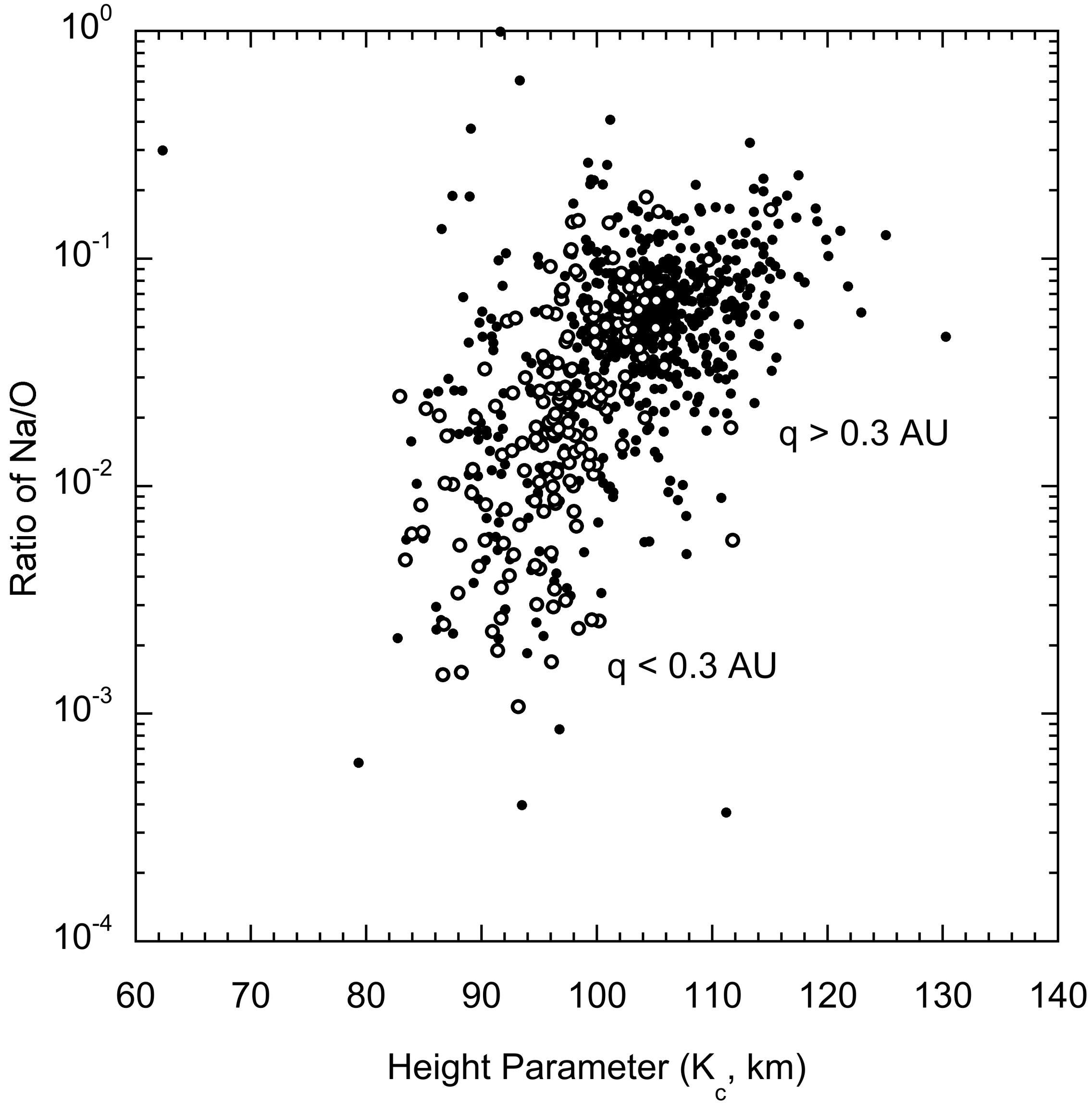

Figure 7B – As Fig. 7A, for all meteoroids with q > 0.3 AU. Open circles mark the meteors from the apex source (generally from long-period comets) that have Type III beginning heights.

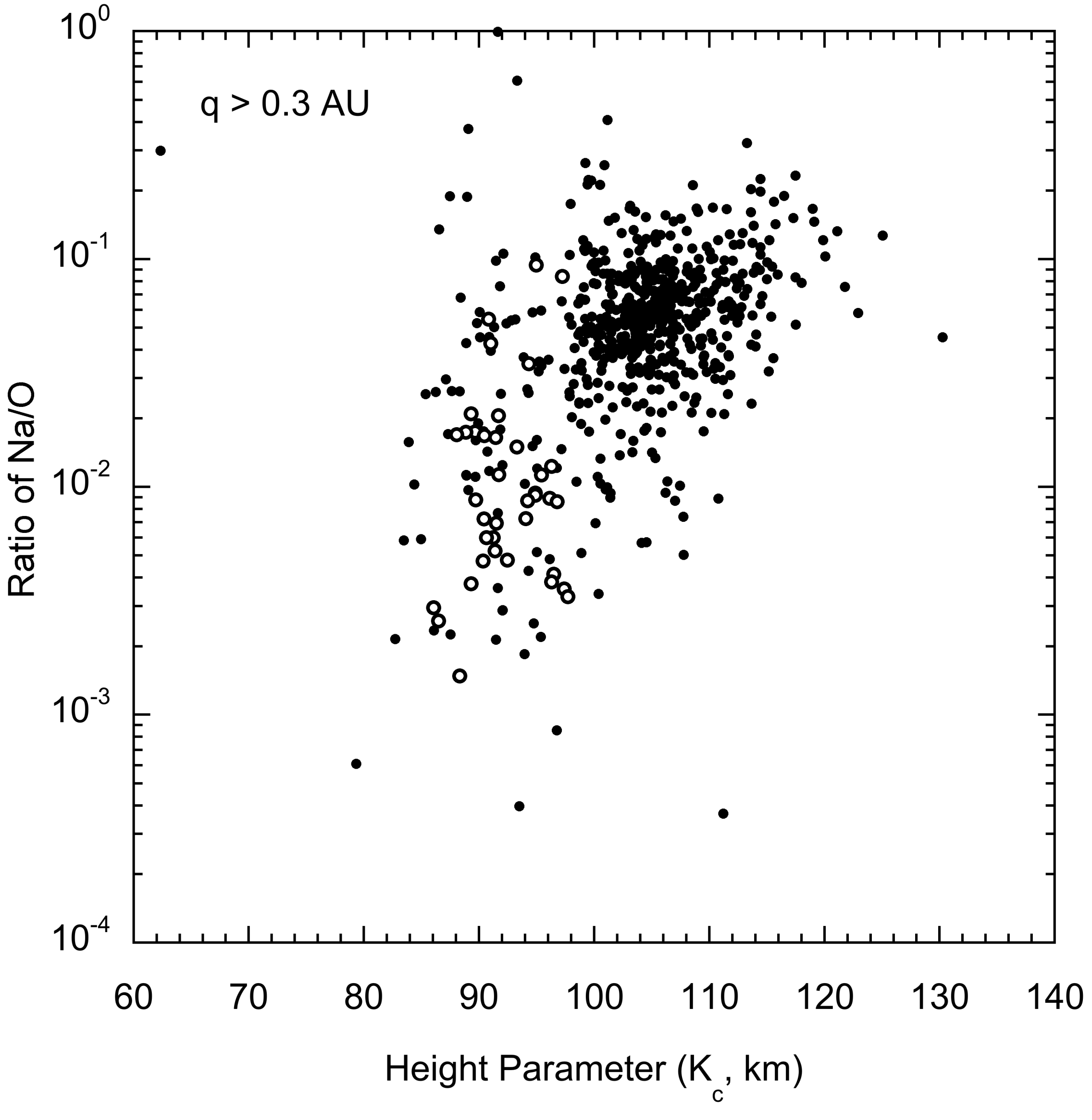

Figure 8 – Meteoroid density (ρ) as a function of the median semi-major axis of the current orbit for all video-detected meteor showers, shown separately for height class Type I (filled circles) and height class Type III (gray diamonds) meteors. Points with open squares and circles are the young and episodic showers. Jupiter-family comet showers are left of the dashed line and labeled "Short Period".

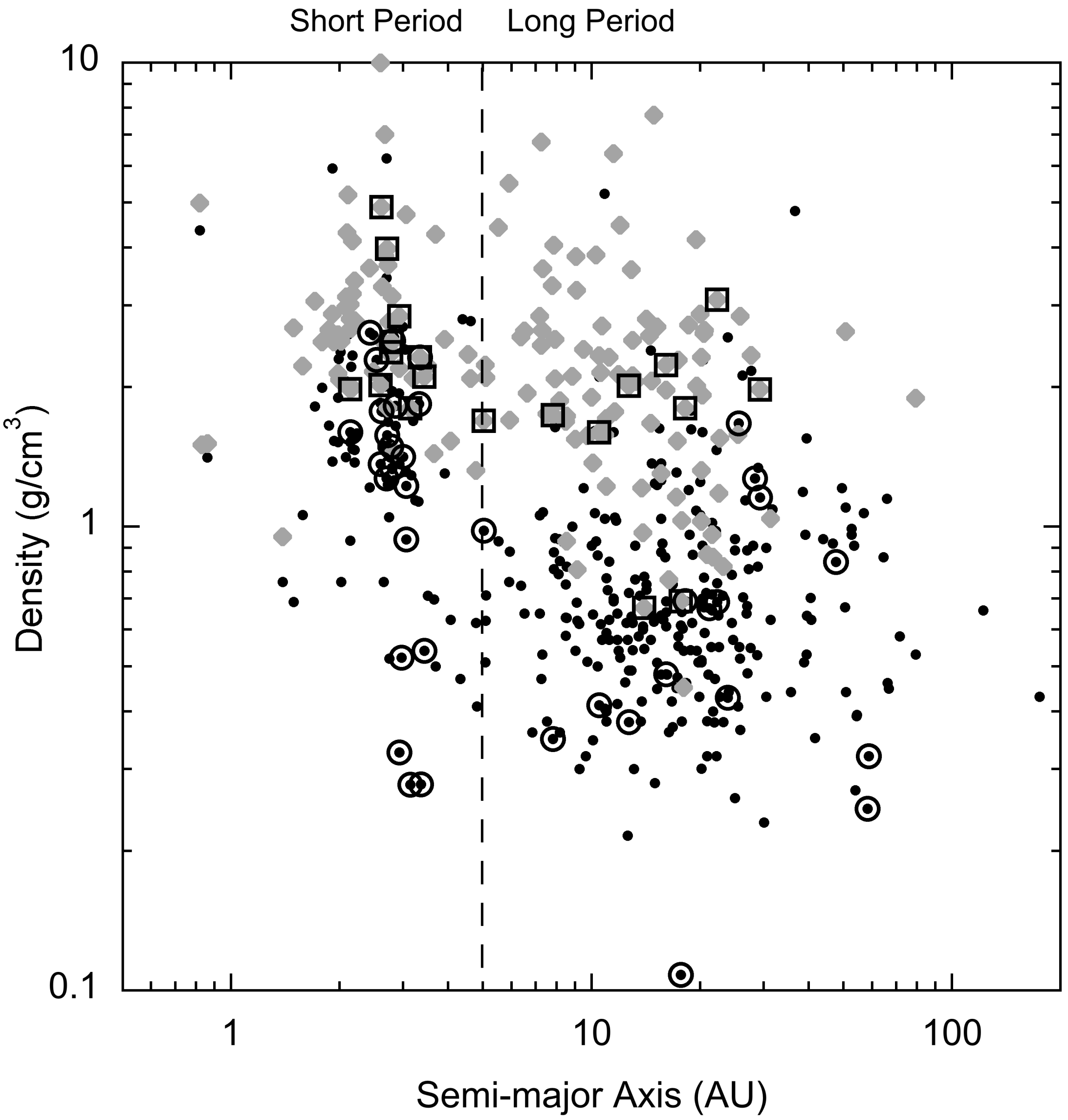

Figure 9 – Magnitude distribution index (χ) as a function of semi-major axis of the current median orbit for all young and episodic showers (with long-period comet showers marked as filled circles, while Jupiter-family comet showers are open circles, both with error bars), and for all long-period comet showers that have χ corrected for age to the time of ejection (+).

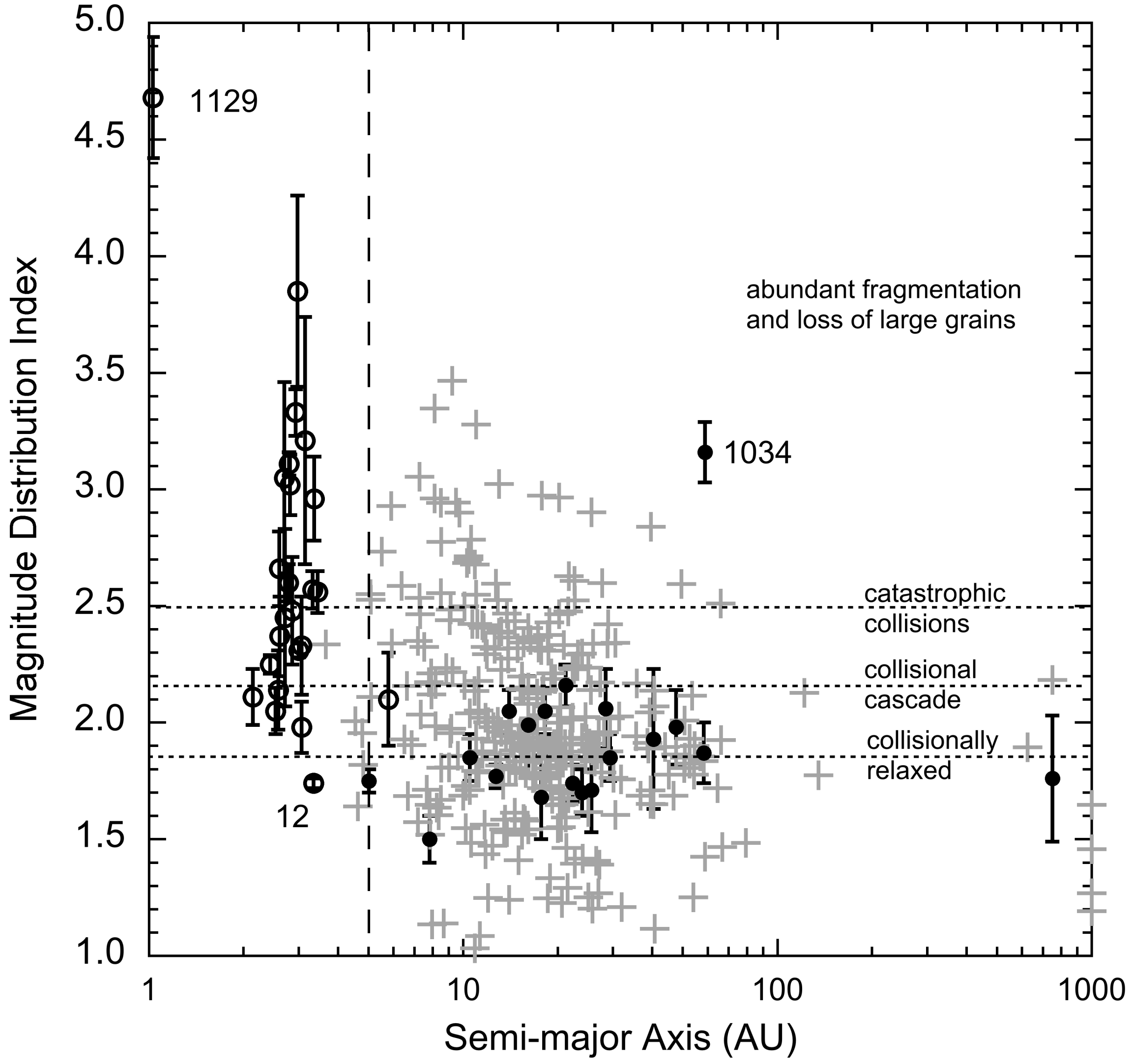

Figure 10 – Magnitude distribution index as a function of the meteoroid density, restricted to meteor showers with q > 0.3 AU. Long-period comet showers are marked as filled circles, while Jupiter-family comet showers are open circles. Long-period comet showers that have $\chi$ corrected for age to the time of ejection are given as gray crosses. Draconids (shower #9), Andromedids (#18), and $\gamma$-Piscis Austrinids (#1034) are labeled.

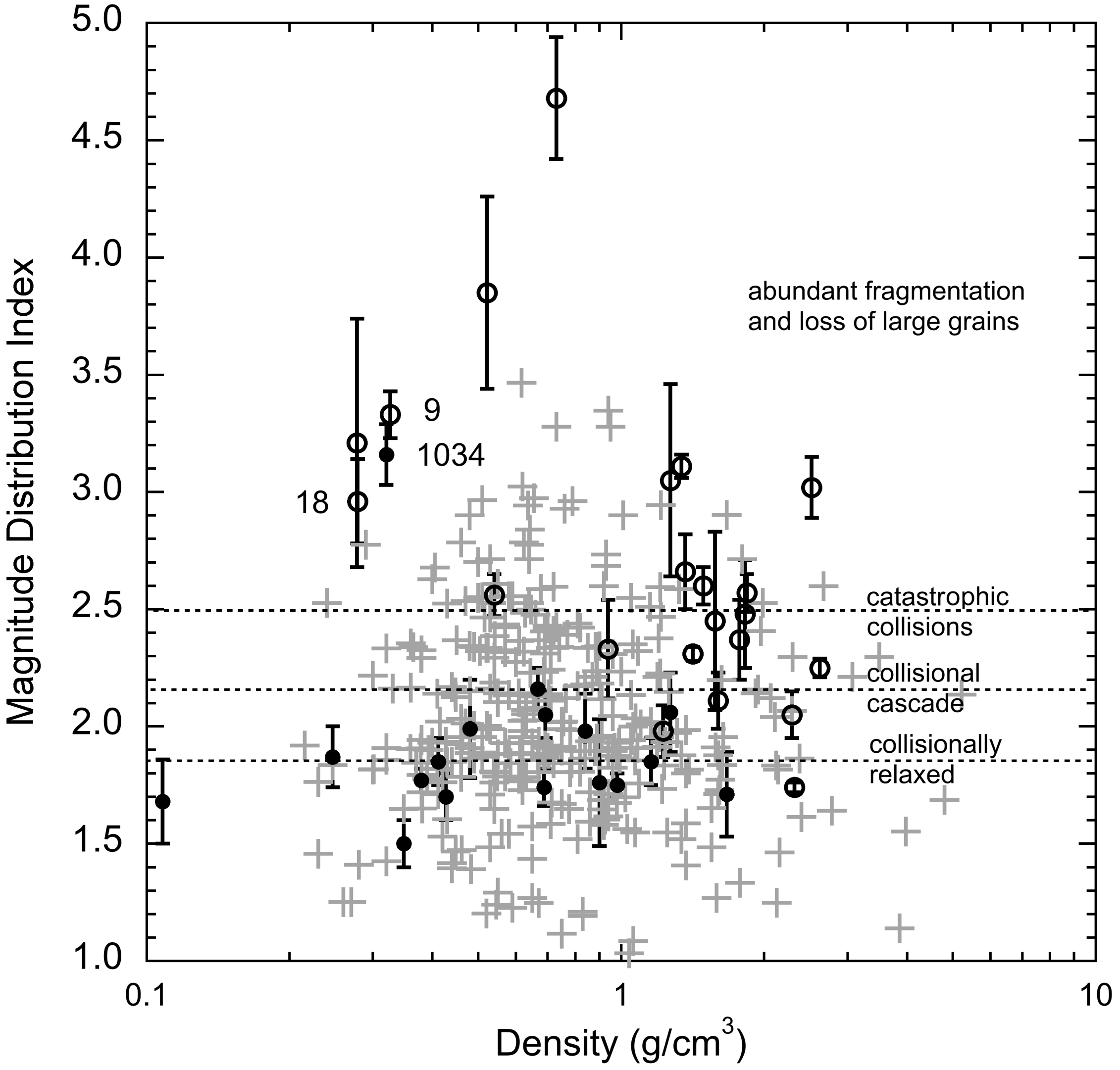

Figure 11 – The magnitude distribution index as a function of the light curve shape parameter F for all showers with q > 0.3 AU. Symbols as before: open circles are young JFC type showers, while filled circles are young LPC type showers and crosses are LPC with $\chi$ adjusted for age.

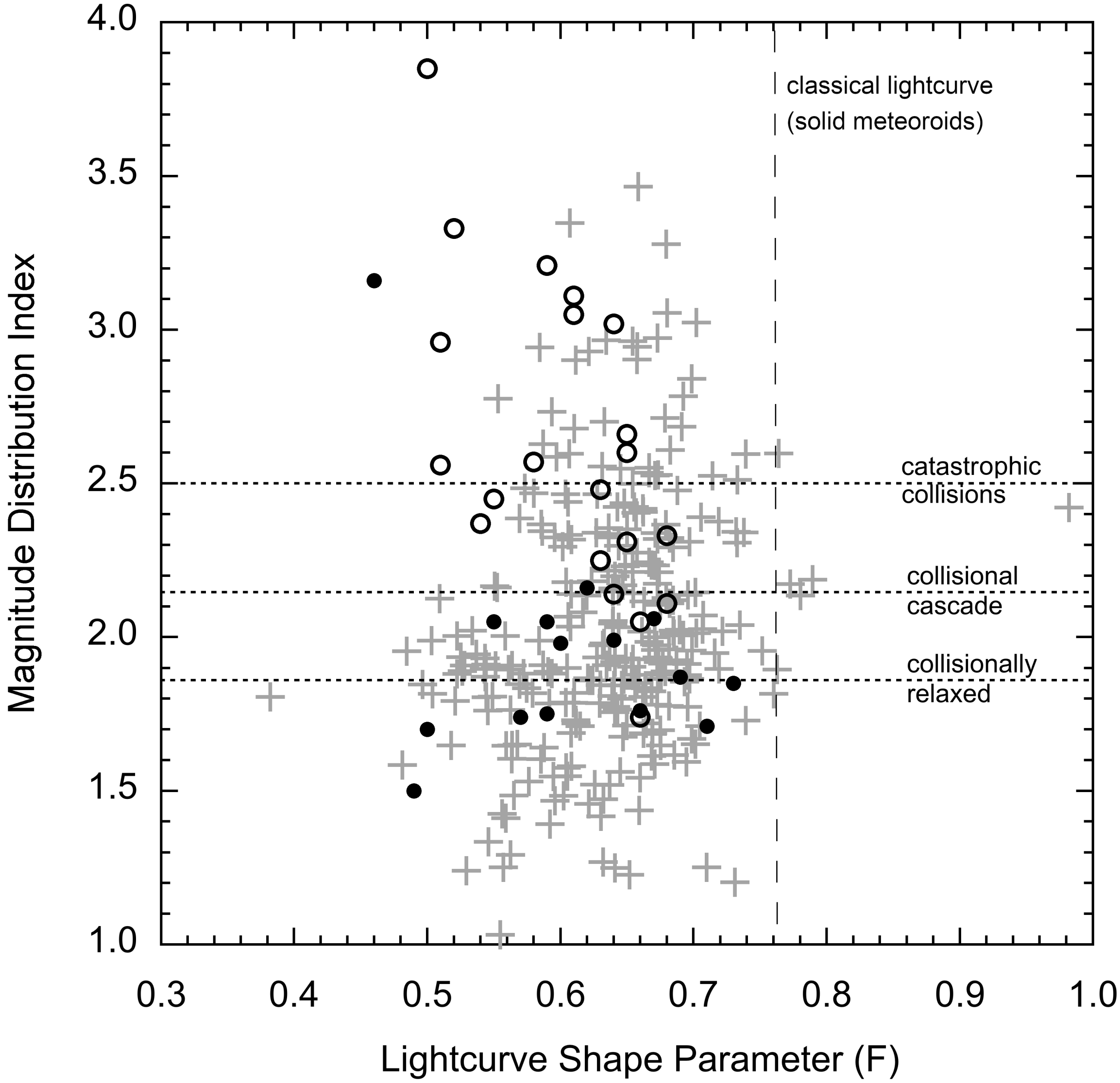

Figure 12A – The admixture fraction (number ratio of height class Type III over all materials) as a function of semi-major axis of the orbit, for all showers with perihelion distance > 0.3 AU and at least 100 detected meteors. Filled circles are young LPC type showers and open circles are young JFC type showers, while gray crosses are all types of showers with $\chi$ adjusted for age.

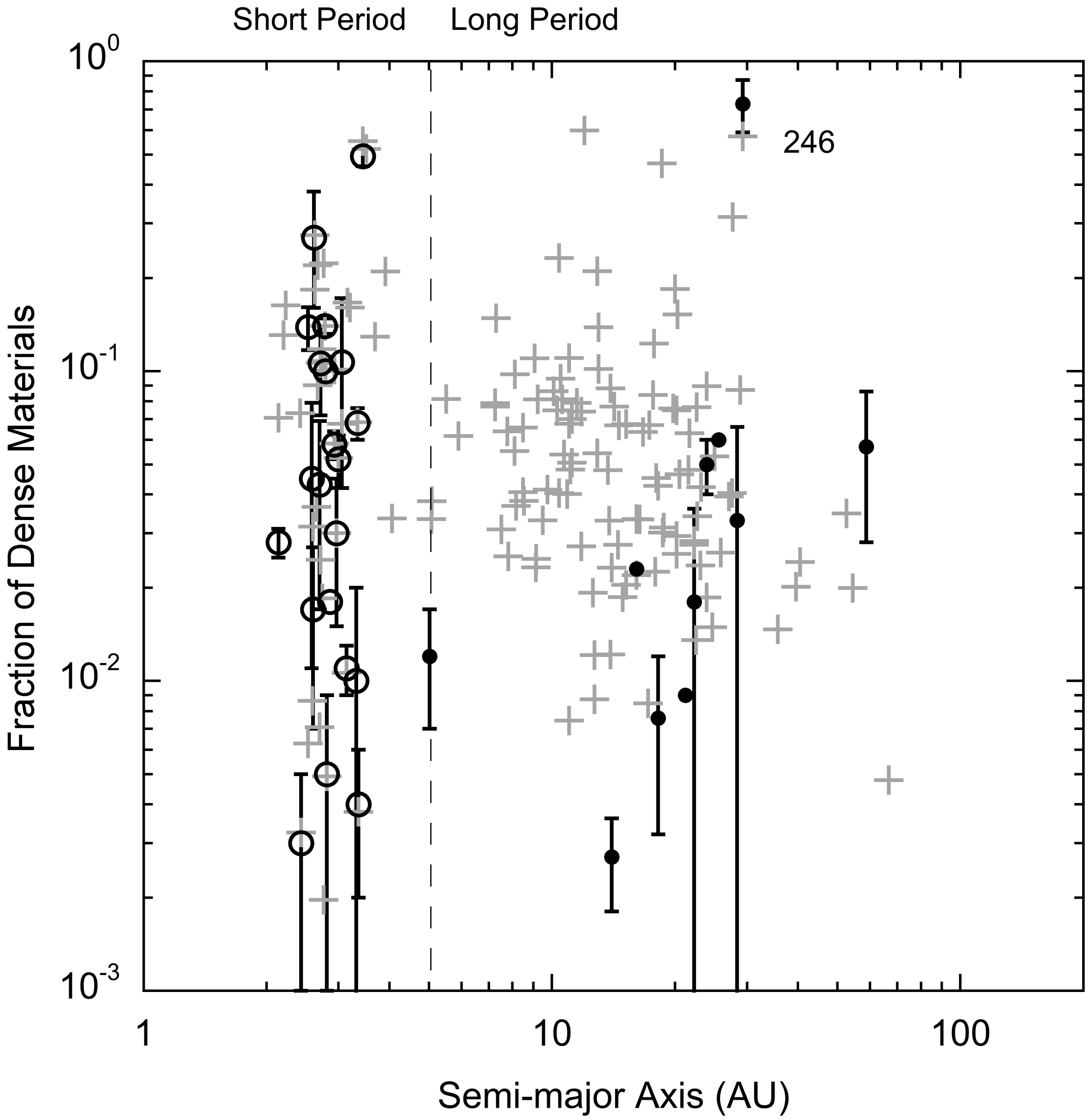

Figure 12B – Admixture fraction of dense materials as a function of inclination of the orbit. Filled circles are young LPC showers, open circles are young JFC type showers. Now gray crosses "+" are LPC with $\chi$ adjusted for age, while gray symbols "x" are JFC ("short period" comets) with $\chi$ adjusted for age.

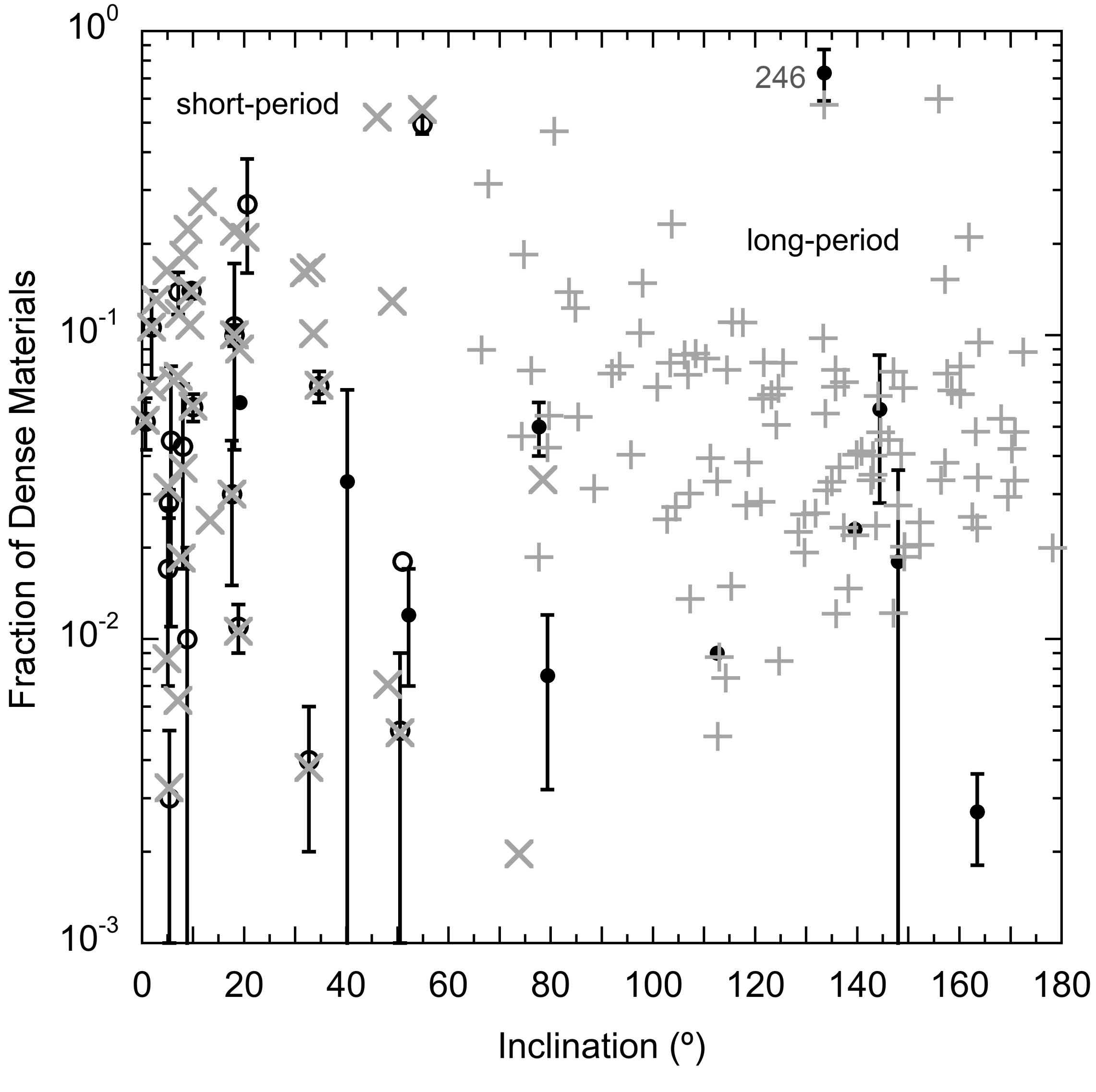

Figure 13 – Evolution of the osculating perihelion distance (q) at each perihelion for LPC type meteor showers over the typical age of such observed streams, evolving from comets with a starting period of 4000 years (left panel) or 250 years (right panel). Symbols change from stars to crosses when the period falls under 200 years, where mean-motion resonances become important. Typically, the perihelion distance does not change significantly, except when the orbital period falls below 200 years and the value of q changes abruptly. Shower orbital elements are for high-q group, from high-inclination (retrograde) to low-i (prograde): #519: q = 0.92 AU, i = 156 º; #1047: q = 0.93 AU, i = 103º; #1146: q = 0.91 AU, i = 74º; #130: q = 0.96 AU, i = 60º; (medium-q group) #206: q = 0.68 AU, i = 148º; #569: q = 0.68 AU, i = 114º; #647: q = 0.64 AU, i = 22º; (low-q group) #16: q = 0.26 AU, i = 129º; and #331: q = 0.31 AU, i = 62º.

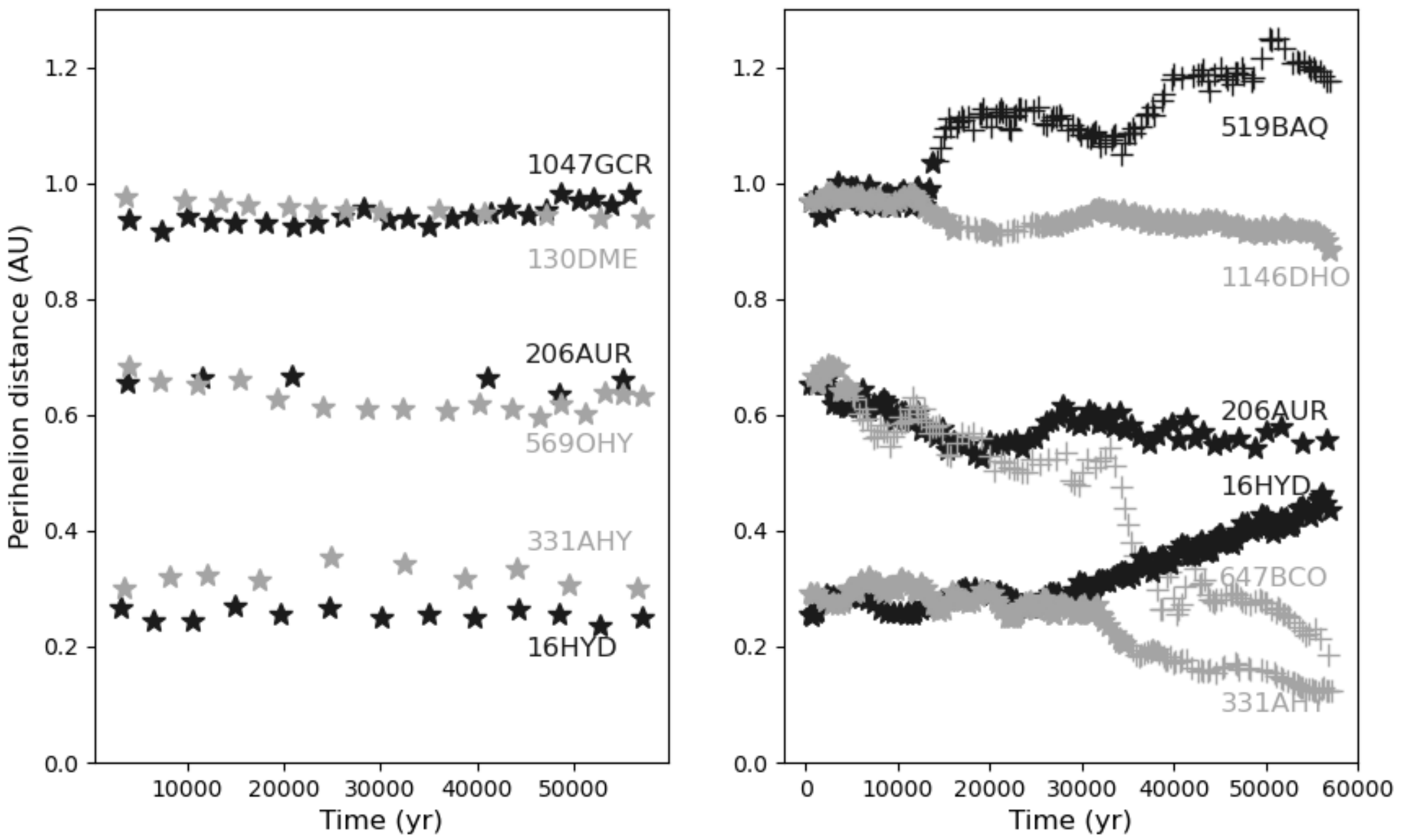

Figure 14 – A schematic diagram depicting early solar system evolution and indicating the possible source regions of future primitive asteroids (AST), Jupiter-family comets (JFC) and long-period comets (LPC), with planet positions after the preferred model discussed in Morbidelli & Nesvorny (2020); (A) Initial protoplanetary disk at ~3 Ma, when gas is about to dissipate and planetesimals are formed; (B) Just before planetary instability at ~50 Ma, when planet migration has created a massive Scattered Disk and Neptune has evolved to 27.7 AU; (C) Situation after planet instability and after Neptune has accreted and scattered the final bit of the Trans Neptunian Disk, with Oort Cloud (about 10,000 – 100,0000 AU from the Sun) drawn not to scale.

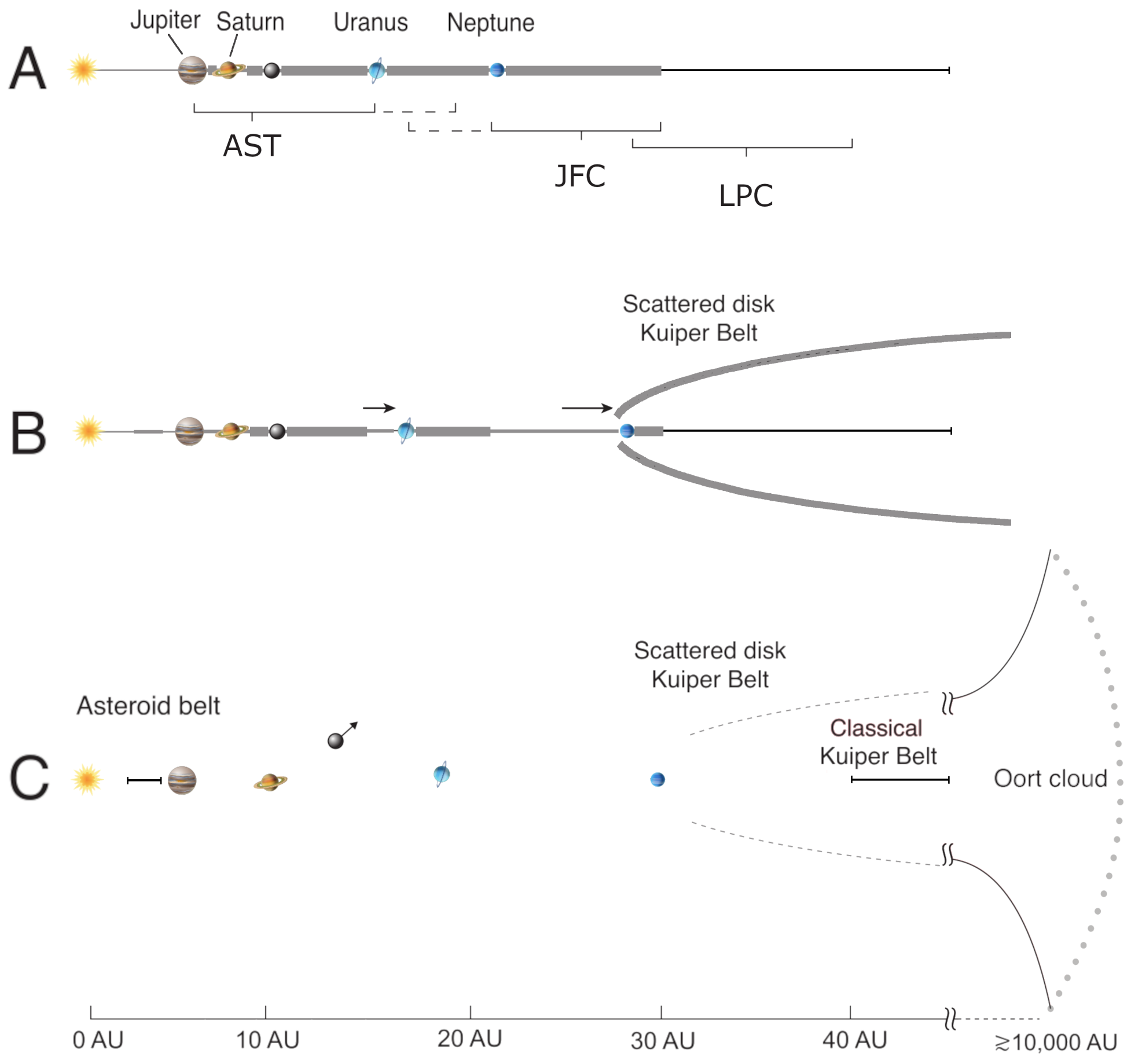